\documentclass[a4paper,12pt]{article}

\usepackage{amsmath,amssymb,amsfonts}
\usepackage{epsfig,cancel,cite}

\pdfoutput=1

\newlength{\xtrawidth}
\newlength{\xtraheight}
\newcommand{\cref}[1]{Chapter~\ref{#1}}
\newcommand{\bcenter}{\begin{center}}
\newcommand{\ecenter}{\end{center}}
\newcommand{\beq}{\begin{equation}}
\newcommand{\eeq}{\end{equation}}
\newcommand{\bea}{\begin{eqnarray}}
\newcommand{\eea}{\end{eqnarray}}
\newcommand{\bean}{\begin{eqnarray*}}
\newcommand{\eean}{\end{eqnarray*}}
\newcommand{\ba}{\begin{array}}
\newcommand{\ea}{\end{array}}
\newcommand{\ben}{\begin{enumerate}}
\newcommand{\een}{\end{enumerate}}
\newcommand{\bi}{\begin{itemize}}
\newcommand{\ei}{\end{itemize}}
\newcommand{\bd}{\begin{description}}
\newcommand{\ed}{\end{description}}
\newcommand{\bdiag}{\begin{diagram}}
\newcommand{\ediag}{\end{diagram}}
\def\fnote#1#2{\begingroup\def\thefootnote{#1}\footnote{#2}
     \addtocounter{footnote}{-1}\endgroup}

\def\IC{\mathbb{C}}

\def\IR{\mathbb{R}}
\def\IP{\mathbb{P}}
\def\IZ{\mathbb{Z}}
\def\cO{{\mathcal O}}

\newcommand{\nn}{\nonumber}

\newcommand{\CA}{{\cal A}}

\newcommand\CN{{\cal N}}
\newcommand\CO{{\cal O}}

\newcommand{\be}{\begin{equation}}
\newcommand{\ee}{\end{equation}}

\newcommand{\comment}[1]{}

\newtheorem{theorem}{\sf THEOREM}

\def\IC{\mathbb{C}}

\def\IR{\mathbb{R}}
\def\IP{\mathbb{P}}
\def\IZ{\mathbb{Z}}

\begin{document}


\title{\LARGE \bf{Heterotic Models from Vector Bundles on Toric Calabi-Yau Manifolds}}
\author{
Yang-Hui He${}^{1,2,3}$,
Seung-Joo Lee${}^{1}$,
Andr\'e Lukas${}^{1}$
}
\date{}
\maketitle
\begin{center}
{\small
${}^1${\it Rudolf Peierls Centre for Theoretical Physics, Oxford
  University,\\
$~$ 1 Keble Road, Oxford, OX1 3NP, U.K.}\\[0.2cm]
${}^2${\it Merton College, Oxford, OX1 4JD, U.K.}\\[0.2cm]
${}^3${\it Department of Mathematics, City University London,\\ 
Northampton Square, London, EC1V 0HB, U.K.}\\
\fnote{}{hey@maths.ox.ac.uk}
\fnote{}{s.lee1@physics.ox.ac.uk}
\fnote{}{lukas@physics.ox.ac.uk}
}
\end{center}

\abstract{
We systematically approach the construction of heterotic $E_8 \times E_8$ Calabi-Yau models, based on compact Calabi-Yau three-folds arising from toric geometry and vector bundles on these manifolds. 
We focus on a simple class of 101 such three-folds with smooth ambient spaces, on which we perform an exhaustive scan and find all positive monad bundles with $SU(N)$, $N=3,4,5$ structure groups, subject to the heterotic anomaly cancellation constraint.
We find that anomaly-free positive monads exist on only 11 of these toric three-folds with a total number of bundles of about 2000. 
Only 21 of these models, all of them on three-folds realizable as hypersurfaces in products of projective spaces, allow for three families of quarks and leptons. 
We also perform a preliminary scan over the much larger class of semi-positive monads which leads to about 44000 bundles with 280 of them satisfying the three-family constraint. 
These 280 models provide a starting point for heterotic model building based on toric three-folds. 
}

\newpage

\tableofcontents

\newpage
%
\section{Introduction}
Heterotic compactification has recently been met with some renewed interest and substantial development.
This most traditional method of string phenomenology involves a succinct formalism in terms of stable holomorphic vector bundles on smooth, compact Calabi-Yau manifolds. Current progress is mainly due to advances in algebraic geometry, both conceptual and computational, the latter facilitated greatly by the ever-increasing power of computers and new algorithms.
In particular, a programme has been established over the past few years on the systematic investigation of the so-called ``general embedding'' realised by special unitary bundles of ranks 3, 4 and 5, on large datasets of the Calabi-Yau threefolds \cite{cyclic, proj, Gabella:2008id, Anderson:2009ge}.
Specifically, extensive use has been made of the ``monad construction'', one of the most efficient methods in creating vector bundles on projective varieties \cite{monadbook}.
Such a construction has been utilised throughout the years in string model building \cite{Distler:1987ee, Kachru:1995em, Blumenhagen:1997vt, Douglas:2004yv}.
A database of monad bundles was constructed in Ref.~\cite{proj}, based on complete intersection threefolds in products of projective spaces, or CICYs, a famous set of 7890 three-folds first classified in Ref.~\cite{Candelas:1987kf, grouporder, Green:1987cr, He:1990pg, Gagnon:1994ek}.
On these, a total of 7118 positive bundles were found and the associated particle content and interactions, computed.
The result was conducive to an algorithmic approach to string phenomenology, making possible the construction of a plethora of candidate models and the systematic selection of promising GUT or standard-model like theories.

It is expedient to summarise here the key features of heterotic compactification in our context which will be of use later.
For a more complete discussion see for example \cite{Candelas:1985en, Witten:1985bz, GSW, Donagi:2004ia}.
\begin{itemize}
\item An $SU(N)$ stable holomorphic vector bundle $V$ on a Calabi-Yau threefold $X$ breaks the $E_8$ gauge theory down to an ${\cal N}=1$ four-dimensional GUT theory with gauge group $E_6$, $SO(10)$ and $SU(5)$, respectively for $N=3,4,5$.
\item The first Chern class of the bundle vanishes: $c_1(V) = 0$.
\item The second Chern class of $V$, $c_2(V)$, is constrained by the second Chern class $c_2(X)$ of the manifold $X$ through Green-Schwarz anomaly cancellation.
\item The number of families and anti-families is given by the dimensions of the bundle cohomologies $H^1(X,V)$ and $H^2(X,V)$, respectively.
\item Stability of $V$ implies that the cohomology groups $H^0(X,V)$ and $H^3(X,V)$ vanish, and, hence, the Atiyah-Singer index theorem shows that the index
$\mbox{ind}(V) = \frac{1}{2} \int_X c_3(V) = -h^1(X,V) + h^2(X,V)$ provides the net number of generations.
\end{itemize}
To break the $SU(N)$ group further one requires a non-trivial first fundamental group of the three-fold and a Wilson line. The former is usually achieved by identifying a freely acting discrete symmetry $G$ of $X$ ``upstairs'' and forming the ``downstairs'' quotient $\tilde{X}=X/G$. In addition, the bundle $V$ on $X$ needs to descend to a bundle $\tilde{V}$ on $\tilde{X}$, typically a non-trivial constraint. 
Here, we will not address this aspect of the construction in detail but merely impose a necessary condition for such a ``downstairs'' model to exist and to produce three families. The ``upstairs'' and ``downstairs'' indices are related by ${\rm ind}(\tilde{V})={\rm ind}(V)/k$, where $k=|G|$ is the order of the discrete symmetry group. We will require three families ``downstairs'', that is ${\rm ind}(\tilde{V})=3$, and that $k$ divides $\chi (X)$, the Euler number of the three-fold, a necessary condition for the existence of a free quotient. In addition, we will use the more refined topological invariants of $X$ introduced in Ref.~\cite{grouporder} in order to further constrain the group order $k$.

In the present paper, we will take the first steps to carry out the aforementioned programme for the largest available class of Calabi-Yau three-folds available, namely the hypersurfaces in toric varieties classified in Refs.~\cite{k:6, reflexive, k:5, k:4, k:2, k:3, k:1} and consisting of some 500 million manifolds. From those manifolds, 124 embed into smooth toric ambient spaces and 101 of those have a particularly simple structure of their K\"ahler cone (the number of K\"ahler cone generators equals $h^{1,1}(X)$). In this paper, we will make a modest start and focus on these 101 toric manifolds and the bundles which can be constructed on them. For brevity, we henceforth refer to  ``Calabi-Yau hypersurfaces in a toric variety'' as ``toric Calabi-Yau manifolds''.\footnote{Of course, this is a slight abuse of nomenclature, since there are no compact, toric varieties which are Calabi-Yau (see, for instance, Ref.~\cite{bouchard}).} We hope that methods similar to the one developed for this relatively small set can ultimately be applied to a very large class of manifolds and bundles in a systematic search for the standard model from heterotic Calabi-Yau compactifications. 
~\\

The organization of the paper is as follows.
In Section 2 we collect the relevant facts on constructing smooth Calabi-Yau threefolds as hypersurfaces in an ambient toric fourfold, focusing especially on the 101 manifolds of interest; we leave some more detailed discussion to the appendices.
In Section 3 we show how to construct monad bundles on these toric hypersurfaces, and how constraints on the Chern classes come from various mathematical and physical restrictions.
We proceed to show that a large class, the so-called ``positive'' monads are finite in number and in Section 4 present their complete classification. In Section 5 we extend our search to semi-positive monads and we conclude with discussion and prospects in Section 6.

\section{The Base Manifolds: Calabi-Yau Threefolds as Hypersurfaces in Toric Fourfolds}
As mentioned above, the largest known data-set to date of smooth, compact Calabi-Yau threefolds consists of hypersurfaces in ambient toric four-folds and has been constructed in Ref.~\cite{reflexive, k:5}. 
These hypersurfaces are defined by the zero set of a single equation in an ambient toric four-fold $\CA$.
Already, this leads to a substantial number of manifolds, namely $473,800,776$. In this paper, we will focus on the cases where the ambient $\CA$ is, in addition, smooth. It is the purpose of this section to briefly summarise the relevant properties of these Calabi-Yau threefolds, on which we shall construct a large class of vector bundles in the ensuing section.
We shall not give a pedagogical introduction to toric geometry and the reader is referred to many excellent texts \cite{fulton,oda,cox:review,bouchard}.
Instead, we leave a somewhat self-contained collection of nomenclature and pertinent facts to Appendix \ref{ap:construction}, and here focus on the geometrical data of the base Calabi-Yau space, as well as of the ambient toric variety, important to the monad construction.
In due course, we shall often draw similarities with the CICY dataset of Calabi-Yau threefolds embedded in products of projective spaces, studied in detail in \cite{grouporder,cyclic,proj, Candelas:1987kf}, of which we have some intuition and familiarity (cf.~also a recent three-generation model found in \cite{Braun:2009qy}).

The first ingredient is the construction of the ambient four-fold $\CA$; this is the analogue of the product of projective spaces for the CICYs.
The power of toric geometry is in using the combinatorics of integer lattices to encode geometrical information.
The ambient space is specified by a {\bf convex integer polytope} $\Delta$ in $\IR^4$ containing the origin.
We can think of this polytope as a collection of vertices (dimension 0), each of which is a 4-vector with integer entries. Each pair of neighbouring vertices defines an edge (dimension 1), each triple a face (dimension 2), and each quadruple, a facet (dimension 3). Alternatively, we could define the polytope by a list of integer inequalities, each of which slices a facet. The polytope is the convex body in $\IR^4$ enclosed by these facets. 
We will only consider those polytopes containing the origin $(0,0,0,0)$ as an interior point. We define the {\bf dual polytope} $\Delta^\circ$ to $\Delta$ as all vectors in $\IR^4$ whose inner product with all interior points of $\Delta$ is greater than or equal to $-1$, that is, 
\be \label{dualp}
\Delta^\circ = \{ \bold{v} \in \IR^4 ~|~ \left\langle \bold{m},\bold{v} \right\rangle \geq -1\;\; \forall \bold{m} \in \Delta \}.
\ee
To this dual polytope we can associate the collection of cones over its faces which, together, form the {\bf normal fan} $\Sigma$.  This normal fan encodes the information necessary to construct the toric ambient space ${\cal A}$ and a brief review of this construction can be found in Appendix~\ref{ap:toric}. It involves associating to each edge of $\Sigma$ a coordinate $x_\rho$. Each cone in $\Sigma$ determines a patch of the toric variety and these patches are glued together in a way determined by how the cones adjoin each other.

Next, we define a Calabi-Yau hypersurface $X$ in $\CA$.
It turns out that this is straight-forward: as long as the polytope is {\bf reflexive} we can define $X$.
The polytope $\Delta$ is called reflexive if the vertices of its dual $\Delta^\circ$ defined by Eq.~\eqref{dualp} are all integer 4-vectors.
Note that in this case, $\Delta^\circ$ is also a reflexive polytope, by symmetry in the definition.
To a reflexive $\Delta$, we can associate a smooth Calabi-Yau threefold $X$ given by the vanishing set of the polynomial
\begin{equation}
\label{deq}
0 = \sum\limits_{\bold{m} \in \Delta} C_{\bold{m}} \prod\limits_{\rho =
1}^k x_\rho^{\langle \bold{m}, \bold{v_\rho} \rangle + 1} \ ,
\end{equation}
where $C_{\bold{m}}$ are numerical coefficients parametrising the complex structure of $X$, $x_{\rho = 1, \ldots, k}$ are the coordinates of $\CA$, and finally, $\bold{v}_{\rho = 1, \ldots, k}$ are the vertices of $\Delta^\circ$, with $k$ being the number of vertices in this dual polytope or equivalently, the number of facets in the original polytope $\Delta$.

As a concrete example, the quintic manifold in $\IP^4$ is a hypersurface in the toric variety $\IP^4$.
We have $x_{1,\ldots,5}$ as the (homogeneous) coordinates of $\IP^4$ and can think of the reflexive polytope $\Delta$ as having vertices
\begin{equation}\begin{array}{rcl}
\bold m_1 &=& (-1,-1,-1,-1), \\
\bold m_2 &=& (~4,-1,-1,-1) , \\
\bold m_3 &=& (-1,~4,-1,-1), \\
\bold m_4 &=& (-1,-1,~4,-1), \\
\bold m_5 &=& (-1,-1,-1,~4) \ ,
\end{array}\end{equation}
as well as all the points interior to these extremal points, including, for example, $(0,0,0,0)$.
The dual polytope $\Delta^\circ$ is easily checked to have vertices
\begin{equation}\label{dualpolyQ}\begin{array}{rcl}
\bold v_1 &=& (1, 0,0,0), \\
\bold v_2 &=& (0,1,0,0), \\
\bold v_3 &=& (0,0,1,0), \\
\bold v_4 &=& (0,0,0,1), \\
\bold v_5 &=& (-1,-1,-1,-1) \ .
\end{array}\end{equation}
Then, according to Eq.~\eqref{deq}, each lattice point $\bold{m} \in \Delta$ contributes a quintic monomial in the coordinates $x_{1,\ldots,5}$ to the defining polynomial.
For example, the origin $\bold{m}=(0,0,0,0)$ gives rise to the monomial $x_1x_2x_3x_4x_5$.
We then sum over these monomials, with arbitrary complex coefficients, giving us a homogeneous quintic polynomial which defines the quintic Calabi-Yau three-fold in $\mathbb{P}^4$.

All complex projective spaces and products thereof are toric varieties.
To anchor ourselves, it is worth mentioning that five of the manifolds we shall subsequently encounter are hypersurfaces in products of projective spaces for which monad bundles have already been analysed in the literature \cite{proj}. These are the five manifolds correspond to the ambient spaces, $\IP^4, ~\IP^1 \times \IP^3, ~\IP^2 \times \IP^2, ~\IP^1 \times \IP^1 \times \IP^2, ~\text{and }\IP^1 \times \IP^1 \times \IP^1 \times \IP^1$.
The first is the quintic mentioned above. It is also interesting to point out that the transpose CICYs \cite{Candelas:2007ac} of these five are the so-called cyclic CICYs, which have been studied in Ref.~\cite{cyclic}.

\subsection{Smooth Ambient Spaces and the Selection of $101$ Spaces}
Half-billion reflexive 4-polytopes $\Delta$ and their associated Calabi-Yau threefolds $X$ represent a formidable dataset. Of these, 124 distinguish themselves in that the ambient four-fold $\CA$ is smooth (we emphasise that all $X$ in the list, even if $\CA$ is singular, are smooth).
These smooth toric 4-folds and the corresponding smooth Calabi-Yau 3-folds form a natural starting point.
In this paper we restrict ourselves even further to the 101 pairs, whose toric 4-folds are not only smooth but also equipped with simplicial K\"ahler cones (we will expound more upon this shortly), and thereon we build vector bundles. 
We will call the spaces with the latter property {\bf simple} manifolds. Focusing on this subset leads to a number of technical simplifications which are helpful in dealing with the bundle construction. A systematic analysis of singular toric varieties and their Calabi-Yau hypersurfaces will be the subject of future work.

We will adhere to the notation of Eq.~\eqref{dualpolyQ} and represent both $\CA$ and $X$ by the vertices of the dual polytope $\Delta^\circ$.
For reference, we present the complete dataset of the 124 smooth ambient toric 4-folds in Appendix \ref{ap:database}; the rows are the integer 4-vectors for the coordinates of the vertices.
Furthermore, for comparision, we have marked numbers 1 (the quintic), 2, 7, 26, 40 with a subscript $P$ because these are precisely the 5 manifolds whose ambient spaces are the products of projective spaces.
It is interesting to notice that our dataset includes 10 ambient spaces of the form $\mathcal A = dP_{k_1} \times dP_{k_2}$ and 4 of the form $\mathcal A = dP_{k_1} \times \IP^1 \times \IP^1$ $(k_1,k_2=0,1,2,3)$,  where $dP_k$ is del Pezzo surface with $k$ general points blown-up. 
Table~\ref{tb8} in Appendix \ref{ap:database} lists these ambients separately.
We have also marked 23 numbers with a subscript $N$, which means that their K\"ahler cones are non-simplicial and we did not attempt to analyse them in this paper.

\subsection{Geometrical Data}\label{geomdata}
Armed with our dataset, we now proceed to discuss some geometrical quantities which will be important to the construction of vector bundles on $X$.
Again, we leave the details to Appendix \ref{ap:properties} and will walk the reader through a detailed example in Appendix \ref{ap:database}.

First, we can compute the {\bf Hodge numbers} of $X$ by simple combinatorics \cite{batyrev} of $\Delta^\circ$ (beautifully reflecting mirror symmetry); the relevant equations are explicitly presented in \eqref{hodge11} and \eqref{hodge21}.
It turns out that the equality $h^{1,1}(X) = h^{1,1}(\CA)$ holds for each of the 101 Calabi-Yau 3-folds, which means that all the closed (1,1)-forms of $X$ descend from $\CA$.
We will say that $X$ is \textit{favourable} if it has this property; favourability turns out to be very convenient for the description of line bundles which we will see shortly.
Indeed, for the CICY dataset, containing 7890 threefolds, 4515 of them are favourable in the same sense.
It was on these favourable spaces that monads were classified in Ref.~\cite{proj}.
It is convenient that not only the 101 simple manifolds, but all our 124 manifolds with smooth ambient space are favourable.
To find $h^{1,1}(\CA)$, we use the relation
\begin{equation}
\label{picard}
\text{Pic}(\CA) \simeq H^2(\CA , \IZ) \simeq\IZ^{k-n} \ ,
\end{equation}
where $\text{Pic}(\CA)$ is the Picard group of $\CA$, $k$, as before, is the number of vertices in the dual polytope and $n=\text{dim}_{\IC} \CA = 4$. For favourable manifolds we then have $h^{1,1}(X) = h^{1,1}(\CA)$ and this number can be easily extracted from Table \ref{tb7} in Appendix \ref{ap:database}; one only needs to count the number of vertices and subtract 4 from it. 
For reference, Table \ref{tb1} shows the distribution of Hodge numbers $h^{1,1}(X)$ of the 101 simple manifolds. 

\begin{table}[thb]
{\begin{center}
\begin{tabular}{|c|c|c|c|c|c|c|}\hline 
$h^{1,1}$ & ~1~ & ~2~ & ~3~ & ~4~ & ~5~ & ~6~  \\ \hline \hline
Number & 1 & 9 & 28 & 44 & 18 & 1  \\ \hline 
\end{tabular}
\end{center}}
{\caption{\label{tb1} \sf Number of simple toric Calabi-Yau hypersurfaces $X$ in smooth toric ambient spaces for each value of $h^{1,1}(X)$.
}}
\end{table}

Next, we need a description of the {\bf K\"ahler cone} of $X$.
The K\"{a}hler cone of the toric ambient space $\cal A$ is determined by the structure of its polytope (for the details, see Theorem \ref{cpl} and Theorem \ref{kahler} in Appendix \ref{ap:properties}).
Since our Calabi-Yau hypersurface $X$ is taken to be favourable, every closed $(1,1)$-form in $X$ can be thought of as the pull-back of a $(1,1)$-form in $\cal A$.
Hence, the K\"{a}hler cone of $X$ must contain that of $\cal A$ (note the reverse inclusion).
It is reasonable to suppose that the  K\"{a}hler cone of $X$ is the same as that of $\CA$. 
To be more precise, we first introduce a basis $\{J_r\}$ of $(1,1)$ forms. We will explain the precise definition of this basis shortly.  A general $(1,1)$ form $J$ can then be expanded as $J=t^rJ_r$. We can represent the K\"{a}hler cone of $\cal A$ (and of $X$) by an $m \times h^{1,1}$ matrix $K=\left[ K^{\bar{r}}_{~r}\right]$, such that all $t^r$ satisfying
\begin{equation}
 K^{\bar{r}}_{~r} t^r \geq 0 \mbox{ for } \bar{r}=1, \cdots , m \label{kcdef}
\end{equation} 
correspond to allowed K\"ahler parameters.
Here, the barred index $\bar{r}$ runs over the facets of the K\"ahler cone and $m$ represents the number of these facets.
Since the number of facets cannot be less than the dimension of the cone, we have
\be \label{simple}
m \geq h^{1,1} \ .
\ee
Our definition of {\it simpleness}, for our database of 101 Calabi-Yau threefolds, is then when \eqref{simple} is saturated, that is, $m=h^{1,1}$.
Appendix~\ref{ap:properties} explains in detail how the matrix $K$ can be determined.

We will also need the {\bf Mori cone} of effective curve classes on $X$; this will be crucial to check the anomaly cancellation conditions. Mori cone is the dual cone to the K\"ahler cone and can thus be determined from the latter readily.

Furthermore, we will require the {\bf Chern classes} and the {\bf intersection numbers} of $X$; these can be determined by a restriction from $\CA$. Indeed, the Adjunction formula dictates that we have the following relation
\begin{equation}
c(\mathcal A) = c(X) \wedge c(\mathcal N) \label{cAX}
\end{equation}
between the total Chern classes of $\CA$ and $X$, where $\CN$ is the normal bundle of $X$, of which we have a good understanding because its Chern class is simply the (multi-)degree of the defining polynomial of $X$ in $\CA$. In practice, these degrees can be obtained from the so-called {\bf charge matrix} $\beta^r_{~\rho}$ which follows from the linear relations between the vertices $\bold v_{\rho=1, \cdots, k}$, as described in Appendix \ref{ap:properties}. Given the charge matrix we simply have
\begin{equation}
c_1(\CN) = n^r J_r \ , \;\; \mbox{where }\;\; n^r=\sum\limits_{\rho=1}^{k} \beta^r_{~\rho}\; .
\end{equation}
The Chern class $c(\mathcal A)$ can be again determined by the combinatorics of the toric data and is presented in Appendix \ref{ap:properties} (see \eqref{chern1} and \eqref{chern2}, for the formula).
Using the relation~\eqref{cAX} we subsequently find, apart from the vanishing $c_1(X)$, that:
\bea
c_2(X) &=& \left[ \sum\limits_{1\leq \rho < \sigma \leq k}^{} {\beta^r_{~\rho} \beta^s_{~\sigma} }\right] J_r \wedge J_s \ ,\\
c_3(X) &=& \left[ \sum\limits_{1\leq \rho < \sigma < \tau \leq k}^{} {\beta^r_{~\rho} \beta^s_{~\sigma} \beta^t_{~\tau}}  - (\sum\limits_{1\leq \rho < \sigma \leq k}^{} {\beta^r_{~\rho} \beta^s_{~\sigma}} ) \cdot (\sum\limits_{1\leq \tau \leq k}^{} \beta^t_{~\tau}) \right] J_r \wedge J_s \wedge J_t \ .
\eea

Finally, the intersection numbers on $\CA$ are
\begin{equation} \label{def_inter}
d_{rstu}= \int_{\cal A} J_r \wedge J_s \wedge J_t \wedge J_u \ ;
\end{equation}
note that we slightly abuse notation and refer to both the $(1,1)$-forms in $\CA$ and $X$ as $J^r$ with $r = 1,\ldots,h^{1,1}(X) = h^{1,1}(\CA)$, because all our $X$ are favourable. A number of linear relations for the intersection numbers of $\CA$, explicitly given in Eq.~\eqref{inter}, can be extracted from the toric data and explicitly solved for $d_{rstu}$. Subsequently, the triple intersection numbers $d_{rst}$ of $X$ can be determined from the intersection numbers on $\CA$ by
\begin{equation}
d_{rst} = \int_X J_r \wedge J_s \wedge J_t =
\int_{\mathcal A} J_r \wedge J_s \wedge J_t  \wedge c_1 (\mathcal N) =
n^u d_{rstu} \ .
\end{equation}

\section{Construction of Vector Bundles}
For heterotic string models, gauge bundles need to be constructed over the Calabi-Yau 3-folds. 
In the preceding section, we have introduced the base Calabi-Yau manifolds as hypersurfaces in toric four-folds.
In this section, our purpose is to construct explicit vector bundles on them.
In particular, we will extend the so-called monad construction which has been applied to the CICY dataset in Ref.~\cite{proj}, and arrive at analogous classification results.
\subsection{Line Bundles}
In our vector-bundle construction, we will make frequent usage of line-bundles; they are the basic building blocks of our gauge bundles. We begin by studying line-bundles on the ambient $\CA$ and then consider their restriction to $X$. 

We have seen earlier that $\text{Pic} (\CA) \simeq \IZ^{k-4} \simeq H^2 (\CA, ~\IZ)$, where $k$ is the number of vertices in the dual polytope for $\CA$. Hence, we can denote line bundles on $\CA$ by $\CO_\CA (\bold a)$ for $\bold a \in \IZ^{k-4}$. With the standard basis $\{{\bf e}_r\}$ of unit normal vectors in $Z^{k-4}$, we can then define a basis $\{J_r\} $ of $(1,1)$-forms by setting
\be
 J_r \equiv c_1 ( \mathcal O_{\CA} ( \bold e_r ) ) \ ,\quad r=1, \cdots , k-4 ~(=h^{1,1}(\CA)) \ .
\ee
Relative to this basis, the first Chern class of an arbitrary line bundle $\CO_\CA (\bold a)$ can be expressed as
\begin{equation}
c_1 (\CO_{\CA} (\bold{a}) ) = a^r J_r \ , 
\end{equation}
where the sum over $r$ is implicit. The restriction of $\CO_{\cal A} (\bold{a})$ to the hypersurface $X$ will be denoted by $\CO_X(\bold{a})$. Favourability of $X$ says that we obtain all line bundles on $X$ in this way.
Positive line bundles on $X$ are those whose first Chern class is in the interior of the K\"{a}hler cone. From Eq.~\eqref{kcdef} this means a line bundle ${\cal O}_X({\bf a})$ is positive iff
\begin{equation}
 K^{\bar{r}}_{~r} a^r > 0 \mbox{ for } \bar{r}=1, \cdots , m\; . \label{lbpos}
\end{equation}
For such positive line bundles the Kodaira vanishing theorem implies that $H^i(X,{\cal O}_X({\bf a}))=0$ for all $i>0$, that is, the zeroth cohomology is the only non-trivial one.

\subsection{The Monad Construction} \label{monadsec}
Having understood the properties of the Calabi-Yau manifolds $X$ and the line bundles on them, we are now ready to apply monad construction in order to create vector bundles over $X$. We can form direct sums of such line bundles and a {\bf monad bundle} is essentially the quotient of two such sums. More precisely, a monad bundle $V$ over $X$ is defined by the following short exact sequence:
\beq\label{monad}
0 \to V  \to B  \stackrel{f}{\to} C \to 0 \,
\eeq
where $B=\bigoplus\limits_{i=1}^{r_B} \CO_X(\bold{b}_i)$, $C=\bigoplus\limits_{j=1}^{r_C} \CO_X(\bold{c}_j)$ are direct sums of line bundles of ranks $r_B$ and $r_C$, respectively. 

From the definition, one can readily compute all relevant Chern classes of the monad bundle $V$:
\bea
\label{c}
\nn {\rm rk}(V) &=& r_B - r_C = N \ , \;\text{with } N=3,4, \text{or }5 \ ,  \\
\nn c_1 (V) &=& \left( \sum_{i=1}^{r_B} b^r_i - \sum_{j=1}^{r_C} c^r_j \right) J_r \ , \\
c_2(V) &=& \frac{1}{2} d_{rst} \left(\sum_{j=1}^{r_C} c^s_j c^t_j - \sum_{i=1}^{r_B} b^s_i b^t_i \right) \nu^r \ , \\
\nn c_3(V) &=& \frac{1}{3} d_{rst} 
   \left(\sum_{i=1}^{r_B} b^r_i b^s_i b^t_i - \sum_{j=1}^{r_C} c^r_j
   c^s_j c^t_j \right) \ ,
\label{chernV}
\eea
where the 4-forms $\nu^r$ furnish the dual basis elements to the K\"ahler cone generatos $J_r$, and satisfy the duality relation:
\be
\int_{X} J_r \wedge \nu^s = \delta_r^s.
\ee



As was discussed in Ref.~\cite{proj}, a number of constraints should be imposed on our monad construction. 
Let us summarise these constraints.

\subsubsection{Mathematical Constraints}
\paragraph{Bundleness: }
It is not a priori obvious that the exact sequence~\eqref{monad} indeed defines a bundle rather than a sheaf in general. However, thanks to the theorem by Fulton and Lazarsfeld~\cite{fl} this is the case provided the map $f:B\rightarrow C$ is sufficiently generic and the bundle $C \otimes B^\star$ is globally generated. One can ensure that both conditions are met by requiring that all the line bundles in $C \otimes B^\star =\bigoplus_{i,j}{\cal O}_X({\bf k}_{ij})$ are positive, that is, the vectors ${\bf k}_{ij}\equiv{\bf c}_i-{\bf b}_i$ should all satisfy Eq.~\eqref{lbpos}. So, explicitly, we  demand that
\begin{equation} 
 K^{\bar{r}}_{~s} k_{ij}^s \geq 0 \;\;\forall \bar{r},i,j\; . \label{bundleness}
\end{equation} 

\paragraph{Non-triviality: }
Suppose we have a monad bundle $V_R$ defined by the short exact sequence
\be \label{monadr}
0 \to V_R  \to B \oplus R \stackrel{f_R}{\to} C \oplus R \to 0\;.
\ee
where $R$ is a sum of line bundles. Comparing Eqs.~\eqref{monadr} and \eqref{monad}, one can see that $V_R$ is actually equivalent to $V$. To remove such equivalent monad bundles we should require that no line bundle is contained in both $B$ and $C$. This means that we can somewhat strengthen the bundleness constraint and require, in addition to Eq.~\eqref{bundleness}, that there exists at least one $\bar{r}$ such that $K^{\bar{r}}_{~r} k_{ij}^r > 0$. 

\paragraph{Positivity:} We will call a monad {\bf positive} if both $B$ and $C$ are sums of positive line bundles. From Eq.~\eqref{lbpos} this means a positive monad is characterised by
\be 
K^{\bar{r}}_{~r} b_i^r > 0\;\;{\forall} \bar{r}, i ~;~~ K^{\bar{r}}_{~r} c_j^r > 0\;\;{\forall} \bar{r}, j \ . \label{positivity}
\ee
Unlike the previous two conditions, positivity is primarily a technical requirement which simplifies many calculations due to Kodaira vanishing being applicable. It also has important physical consequences. For example, consider the long exact cohomology sequence
\begin{equation}
 \begin{array}{lllllllll}
  0&\rightarrow&H^0(X,V)&\rightarrow&H^0(X,B)&\rightarrow&H^0(X,C)&&\\
    &\rightarrow&H^1(X,V)&\rightarrow&H^1(X,B)&\rightarrow&H^1(X,C)&&\\
    &\rightarrow&H^2(X,V)&\rightarrow&H^2(X,B)&\rightarrow&H^2(X,C)&&\\
    &\rightarrow&H^3(X,V)&\rightarrow&H^3(X,B)&\rightarrow&H^3(X,C)&\rightarrow&0 \ .
 \end{array}
 \end{equation}   
Given that $H^i(X,B)=H^i(X,C)=0$ for all $i>0$ it follows immediately that $H^2(X,V)=H^3(X,V)=0$. In particular, positive monads do not have anti-families.
There is also a more tenuous connection between positivity and stability of the bundle $V$. It was shown in Ref.~\cite{cyclic} that all positive monads on cyclic CICYs are stable and, indeed, that all non-positive monads are unstable. The relation is less clear on non-cyclic CICYs but in this case stability has been proven for a large number of positive monads and it is suspected that all positive monads are stable. On the other hand, it is also known that on non-cyclic CICYs positivity is not a necessary condition for stability and some explicit examples of non-positive stable monad bundles are known~\cite{Anderson:2009sw,Anderson:2009nt}.
In the following section, we will focus on positive monads, that is monads satisfying the condition~\eqref{positivity} and work out a complete classification of these bundles. Subsequently, we will slightly relax this condition and also study {\bf semi-positive} monads, that is monads, which, instead of \eqref{positivity}, satisfy:
\be 
K^{\bar{r}}_{~r} b_i^r \geq 0\;\;{\forall} \bar{r}, i ~;~~ K^{\bar{r}}_{~r} c_j^r\geq 0\;\;{\forall} \bar{r}, j  \label{spositivity}
\ee

\subsubsection{Physical Constraints}
In addition to the mathematical constraints above, we should also consider physical ones. 
\paragraph{Correct structure group: }
For the structure group of monad bundles to be either $SU(3)$, $SU(4)$ or $SU(5)$, we first need $N=r_B - r_C = 3, 4$ or $5$. 
In addition, $c_1 (V)$ needs to vanish because the structure group is special unitary.
Therefore, we have that
\beq
\label{Svec}
\sum_{i=1}^{r_B} b^r_i = \sum_{j=1}^{r_B - N} c^r_j \equiv S^r\ , ~~ 
\forall r=1, \cdots , h^{1,1}(X).
\eeq

\paragraph{Anomaly cancellation: }
To ensure that 4-dimensional $\mathcal N =1$ gauge theory is anomaly-free upon compactification, we use the standard Green-Schwarz cancellation method. 
We can further allow the existence of a bulk 5-brane which wraps a holomorphic curve $C$, such that its class $W = [C]$ represents a true complex curve. Hence $W$ should be effective, that is, it should be an element of the Mori cone of $X$. If we take, for simplicity, a trivial hidden bundle, the 5-brane class then becomes
\bea
\label{five-brane}
\nn W &=& c_2(X) - c_2 (V) \\
      &=& \left\{c_{2r} (X) - \frac{1}{2} d_{rst} \left(\sum_{j=1}^{r_C} c^s_j c^t_j - \sum_{i=1}^{r_B} b^s_i b^t_i \right)\right\} \nu^r \\ 
\nn      &\equiv& w_r (\{\bold b_i\}, \{\bold c_j\}) \nu^r \ .
\eea
Note, that the five-brane class $W$ is determined by the coefficients $w_r$ which are functions of the integers $b_i^r$ and $c_j^r$. Hence, for each monad we can compute this five-brane class explicitly and, since we have determined the Mori cone for our base manifolds as discussed earlier, we can check if $W$ is indeed effective. For favourable CICYs the Mori cone is the positive quadrant \footnote{To be precise, the terminology positive ``quadrant'' is only valid in dimension 2 but we adhere to this without ambiguity.} and this check amounts to verifying that all $w_r\geq 0$. Here, the situation is somewhat more complicated since the Mori cone of our toric Calabi-Yau manifolds is not necessarily the positive quadrant in our chosen basis $\nu^r$ of four-forms. We will now explain how to deal with this technical complication.

\subsection{Mori Cones and Basis Change in $H^2(X, \IZ)$}
For a simple space, by definition, the K\"{a}hler cone only has $h^{1,1}$ facets and hence, it also has exactly $h^{1,1}$ generators which we denote by $\tilde{J}_r$. 
So the generators $\tilde{J}_{r}$ can be set as the standard basis elements of the $h^{1,1}$-dimensional vector space by an appropriate linear transformation. 
In other words, upon the linear transformation, the K\"{a}hler cone fits into the positive quadrant.
This is a crucial step for the finiteness arguments in the next section. 

With our new basis elements, an arbitrary closed $(1,1)$-form can be re-expressed as 
\begin{equation}
a^s J_s = a^s \delta^t_s J_t = a^s(K^{-1})^t_{~r} K^r_{~s} J_t = \tilde{a}^r \tilde{J}_r
\ ,
\end{equation}
where $\tilde{a}^r = K^r_{~s} a^s$ and $\tilde{J}_r = J_t (K^{-1})^t_{~r}$. 
Note that we no longer distinguish barred indices from unbarred ones and use the unbarred for both upper and lower indices of $K$  since the K\"{a}hler cone matrices are square for simple spaces.

Let $\tilde{\nu}^r$ be the dual basis elements of $\tilde{J}_r$ such that 
\be
\int_{X} \tilde{J}_r \wedge \tilde{\nu}^s = \delta_r^s
\ee 
and let us rewrite the 5-brane class in terms of the new basis:
\be
\label{w}
W=w_r \nu^r = \tilde{w}_r \tilde{\nu}^r. 
\ee 
It is then straightforward to see that the condition for anomaly cancellation gets translated as follows:
\be
\label{anomaly}
\textit{The 5-brane class $W$ is effective if and only if $\tilde{w}_r ={(K^{-1})^s}_rw_s\geq 0$ for all $r$. \\}
\ee
Here, the matrix $K$ which describes the K\"{a}hler cone of $X$ has been introduced in Section~\ref{geomdata} and the $w_s$ are computed from Eq.~\eqref{five-brane}.

\section{Classification of Positive Monads}
We have now laid the groundwork necessary to address the main purpose of this paper, namely, to initiate the systematic study of monad bundles with structure group $SU(N)$, $N = 3,4,5$ over Calabi-Yau threefold hypersurfaces in four complex dimensional toric ambient varieties.
To begin with, we have first restricted to the 124 smooth ambient spaces which all turn out to be favourable, and thence further to the 101 simple spaces where the number of K\"ahler cone generators equals to the dimension of the cone.
On these spaces, we can very easily define monads, especially positive monads where the entries which determine the sums of line bundles $B$ and $C$ in \eqref{monad} are all strictly positive. Some of the reasons for focusing on this data set of positive monads initially have already been explained: technical advantages in computing bundle cohomology due to Kodaira vanishing, the guaranteed absence of anti-families and the likely stability of positive monad bundles. In this section, we will prove another attractive property which has already been observed in the context of CICYs: Subject to the constraints explained in the previous section positive monad bundles form a finite set. This opens up the possibility of a complete classification which we will carry out explicitly.

\subsection{Finiteness of the Classification Programme}
One obvious question to ask before we start the actual search for positive monads is whether there are finitely many solutions given the constraints described in the previous section. To answer this question, we begin by re-stating the problem in a more formal way. We translate the list of constraints in the previous section to a set of explicit Diophantine (in)equalities, in complete analogy to the CICY case in \cite{proj}.
For any simple Calabi-Yau hypersurface $X$ defined in a nonsingular toric 4-fold, and for any $N=3,4,5$, we wish to find all sets of integers $\tilde{b}_i^r$ and $\tilde{c}_j^r$, where $r=1, \cdots, h^{1,1}(X), i=1, \cdots, r_B = r_C + N \text{ and } j=1,\cdots,r_C$, satisfying the following constraints:
\bea
\label{bundle-constraints}
\nn 1. && \tilde{b}_i^r \geq 1, ~~ \tilde{c}_j^r \geq 1 \, ,\;\;\forall\, i,j,r~; \\
\nn 2. && \tilde{k}_{ij}^r \geq 0 \;\;\forall\, i,j,r \text{ where } {\tilde{k}}_{ij}^r = {\tilde{c}}_j^r - {\tilde{b}}_i^r ~; \\
\nn 3. && {}\forall\, i,j,\;\;\exists\, r ~\text{such that } \tilde{k}_{ij}^r > 0 ~; \\
4. && \sum\limits_{i=1}^{r_B} \tilde{b}_i^r = \sum\limits_{j=1}^{r_C} \tilde{c}_j^r =\tilde{S}^r\, ,\;\;\forall\, r~; \\
\nn 5. && \tilde{d}_{rst} \left(\sum\limits_{j=1}^{r_C} \tilde{c}_j^s \tilde{c}_j^t - \sum\limits_{i=1}^{r_B} \tilde{b}_i^s \tilde{b}_i^t \right) \leq 2 \tilde{c}_{2r}(X)\, ,\;\; \forall\, r\; . 
\eea
Here, tilded quantities are obtained by transforming lower $r,s,t$-type indices of their un-tilded counterparts with ${(K^{-1})^{s}}_r$ and upper indices with ${K^s}_r$, so, for example
\bea
\nn \tilde{d}_{rst} &=& d_{r's't'} (K^{-1})^{r'}_{~~r} (K^{-1})^{s'}_{~~s} (K^{-1})^{t'}_{~~t} \ , \\
\nn \tilde{b}^r_i &=& {K^r}_{r'}b^{r'}_i\ . 
\eea 
Here, $K$ is the matrix which describes the K\"ahler cone of the manifold and was introduced in Section~\ref{geomdata}. A few lines of algebra (see Eq.~(5.7) in Ref.~\cite{proj}) then lead us to the following inequality on $\tilde{b}_{max}^r = \text{max}_i \{\tilde{b}_i^r\}$:
\beq \label{bmax}
\frac{2}{N}\tilde{c}_{2r} (X) \geq  M_{rs} \tilde{b}_{max}^s, 
\eeq
where $M_{rs} = \sum\limits_{t=1}^{h^{1,1}} \tilde{d}_{rst}$. 
It turns out that these inequalities provide upper bounds of $\tilde{b}_{max}^r$ for every simple Calabi-Yau 3-fold on which we are working. 
Moreover, since each $\tilde{b}_{max}^r$ is a strictly positive integer, not all of the 101 simple spaces admit solutions to $\tilde{b}_{max}^r$. 
In fact, the inequalities above immediately eliminate all but 18 spaces, which include the 5 products of projective spaces studied in Ref.~\cite{proj}. 

In order to proceed further, having bounded the maximal entries of the bundle $B$, we now find an upper bound of $r_B$, the rank of $B$. 
\comment{
Again by some algebra, we can get three independent ways to get upper bounds. 
Of course, there may be a more brilliant method than these three.
Once we get a bound for $r_B$, however, whether good or poor, that is already enough as long as computers can sort out all the remaining tasks concerning our classification programme. 
The classification has indeed finished and hence, we do not intend to go any further than these three methods. Below we list all the three without proofs, which can be found in \cite{proj}:
}
This once again proceeds along the same lines as Section 5 of Ref.~\cite{proj}.
There turn out to be three independent bounds, and for each Calabi-Yau, we can check which one leads to the strongest constraint, which is then used in any further calculations. These independent constraints are inequalities (5.13), (5.14) and (5.16) of Ref.~\cite{proj}:
\begin{enumerate}
\item
Given the calculated values of $\tilde{b}_{max}^r$, the following inequality gives us an upper bound: 
\beq 
r_B \leq N \left( 1+ \sum\limits_{r=1}^{h^{1,1}} \tilde{b}_{max}^r \right) \ . 
\eeq
\item
We first find non-negative integers $u^r$, satisfying 
\beq 
M_{rs} u^s \leq 2 \tilde{c}_{2r} (X). \label{bd} 
\eeq 
Note that the inequality above has essentially the same form as the one \eqref{bmax} for $\tilde{b}_{max}^r$ and, therefore, the solution space for the $u^r$ is finite. The non-negative integers $u^r$ are related to $r_B$ by 
\be \label{rb}
r_B = N+\sum\limits_{r=1}^{h^{1,1}} u^r  \ .
\ee
Given the finite solution set for $u^r$, we take the maximum of the corresponding $r_B$ values. 
\item
As in method 2, we first solve the inequality below for non-negative integers $u^r$: 
\beq \sum\limits_{s=1}^{h^{1,1}} \left( 2 \sum\limits_{t=1}^{h^{1,1}} \tilde{d}_{rst} \tilde{b}_{max}^t + \tilde{d}_{rss} \right) u^s \leq 2 \tilde{c}_{2r} (X) + N \tilde{d}_{rst} \tilde{b}_{max}^s \tilde{b}_{max}^t. \eeq
Then we calculate all possible values of $r_B$ from Eq.~\eqref{rb} and find their maximum.
\end{enumerate}

Since $r_B$ and $\tilde{b}_{max}^r$ are now both bounded, we conclude that, as in the CICY cases, {\it the number of positive monads over the 101 simple Calabi-Yau hypersurfaces in smooth toric 4-folds is finite, and in fact exists only on 18 of them.}

\subsection{The Classification Results}
Given that our problem is bounded we can now explicitly classify all solutions by a computer scan. 
For each of the 18 simple Calabi-Yaus with solutions to the inequalitiy for $\tilde{{b}}_{max}^r$, we scan over all allowed values of $N, r_B$ and over all values of the sum vector $\tilde{S}^r$. 
This last vector, is again constrained, and is subject to inequality (5.7) of \cite{proj}: 
\beq 
2 \tilde{c}_{2r} (X) \geq \frac{N}{r_B} M_{rs} \tilde{S}^s. 
\eeq
For each fixed set of these quantities we generate all multi-partitions of entries $\tilde{b}_i^r \text{ and } \tilde{c}_j^r$ modulo permutation symmetry, since the order of summands in a direct sum of line bundles is clearly irrelevant. 

\begin{figure}[t]
\centerline{
(a)
\includegraphics[trim=0mm 0mm 0mm 0mm, clip, width=3.0in]{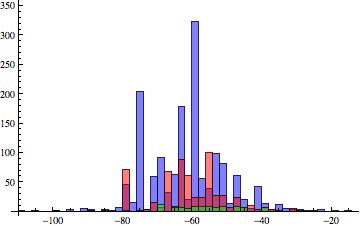}
(b)
\includegraphics[trim=0mm 0mm 0mm 0mm, clip, width=3.0in]{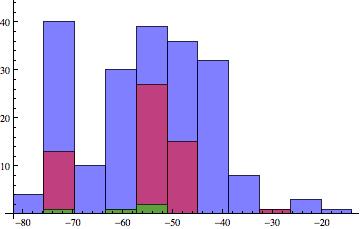}
}
\caption{{\sf The number of positive monads as a function of ${\rm ind}(V)$. Fig.~(a) contains all models, Fig.~(b) only those which satisfy the three-familiy constraint ${\rm ind}(V)=3k$, $k\; |\;\chi(X)$. The three colours blue, red, and green correspond to SU(3), SU(4) and SU(5) models, respectively.}}
\label{f:histo12}
\end{figure}

Upon performing this scan, we find that positive monads only exist over 11 simple Calabi-Yaus out of the 18. 
There are 2190 positive monads in total. The majority of these bundles, namely 1853 of them, arises on the five hypersurfaces in products of projective spaces and is, therefore, already contained in the classification carried out in Ref.~\cite{proj}. The remaining 337 bundles are new. The number of bundles as a function of ${\rm ind}(V)$, the net number of generations, is shown in part (a) of Fig.~\ref{f:histo12} and Table~\ref{tb2} lists the number of solutions for each of the 11 base manifolds. Two explicit examples are
\bea
\nn 1: && 0 \to V^+_1  \to \CO_{X_1} (1,1)^{\oplus 7} \stackrel{f_1}{\to} \CO_{X_1} (5,1) \oplus \CO_{X_1}(1,3)^{\oplus2} \to 0 \ , \\
\nn 2: && 0 \to V^+_2  \to \CO_{X_2} (1,1) ^{\oplus 15} \stackrel{f_2}{\to} \CO_{X_2} (1,2)^{\oplus 5} \oplus \CO_{X_2}(2,1)^{\oplus5}  \to 0 \ ,
\eea
where the first one is an $SU(4)$-bundle on the space number $6$ and the second one an $SU(5)$-bundle on $7_P$ (the numbering of the spaces is according to Table~\ref{tb7} where the toric data for these base spaces can be found). Note that $h^{1,1} = 2$ for both of the spaces.
\begin{table}[thb]
{\begin{center}
\begin{tabular}{|c|c|c|c|c|c|c|c|c|c|c|c|}\hline 
Space No. & $~~1_P~$ & $~~2_P~$ & $~~3~~$ & $~~4~~$ & $~~6~~$ & $~~7_P~$ & $~12~$ & $~17~$ & $~22~$ & $~26_P$ & $~40_P$\\ \hline \hline
SU(3) & 20 & 611 & 4 & 9 & 153 & 38 &  74 & 34 & 9 & 304 & 251 \\ \hline
SU(4) & 14 & 308 & 0 & 0 & 35 & 27 & 0 & 0 & 0 & 135 & 70 \\ \hline
SU(5) & 9 & 56 & 0 & 0 & 19 & 10 & 0 & 0 & 0 & 0 & 0 \\ \hline
\end{tabular}
\end{center}}
{\caption{\label{tb2} \sf Number of positive monad bundles over the 11 CY 3-folds for which positive monads exist. The numbers labelling the space are according to Table~\ref{tb7} where the toric data of the base manifolds can be found. The subscript $P$ indicates that the space is a hypersurface in a product of projective spaces.
}}
\end{table}

\begin{table}[thb]
{\begin{center}
\begin{tabular}{|c|c|c|}\hline 
&  &  \\ 
& $~~~~~\text{No Constraints}~~~~~$ &  $~\text{ind}(V)=3k$, $k ~|~ \chi (X)~$   \\
&  &  \\ \hline \hline
$\text{~~SU(3)~~}$ & 1507 (283) & 204 (59) \\ \hline
SU(4) & 589 (35) & 57 (5) \\ \hline
SU(5) & 94 (19) & 4 (0) \\ \hline \hline
Tot. & 2190 (337) & 265 (64) \\ \hline
\end{tabular}
\end{center}}
{\caption{\label{tb3} \sf Total number of positive monads on the 11 base manifolds (left column) and those which satisfy a basic three-generation constraint (right column). The numbers in the parenthesis only count new monads which have not been already found in Ref.~\cite{proj}.
}}
\end{table}

We would now like to impose a basic three-familiy constraint on our models. We require that the number of families is a multiple of three, that is, ${\rm ind}(V)=3k$ for $k\in\mathbb{Z}_{\neq 0}$, and that the Euler number of $X$ is divisible by the potential group order $k$, that is, $k\;|\;\chi(X)$. These two conditions are clearly necessary (although not sufficient) for the existence of a free quotient $X/G$ with three generations ``downstairs'',  where $|G|=k$. The number of models satisfying these condition is given, as a function of ${\rm ind}(V)$, in part (b) of Fig.~\ref{f:histo12} and their total number is given in Table~\ref{tb3}.

For the above constraints, we have used that possible orders, $k$, of discrete symmetry groups must divide the Euler number of the manifold. There exist a number of more refined topological invariants, given in Ref.~\cite{grouporder}, which can be used to further constrain the group order. These are the Euler characteristics $\chi({\cal N}^k\otimes TX^l)$ and Hirzebruch signatures $\sigma ({\cal N}^k\otimes TX^l)$ of the
``twisted'' bundles ${\cal N}^k\otimes TX^l$ (where ${\cal N}$ is the normal bundle of $X$) which must be
divisible by the group order $|G|$ for all integers $k,l\geq 0$. It was shown in Ref.~\cite{grouporder}, that is it sufficient to consider the cases $(k,l)=(0,1),(1,0),(2,0),(3,0)$ for the Euler characteristic and $(k,l)=(1,1)$ for the Hirzebruch signature without loosing information. We have computed these indices for all the 11 spaces with positive monad bundles, using the equations provided in Ref.~\cite{grouporder}. Their common divisors in any one case provides us with a list, $S(X)$, which must include the orders of all freely-acting symmetry groups for X.
Requiring that $k={\rm ind}(V)/3$ is an element of this list dramatically reduces the number of solutions and we remain with 21 positive monads over 3 Calabi-Yau spaces, all of which are hypersurfaces in products of projective spaces. These 21 models have already been found in Ref.~\cite{proj} and will, therefore, not be discussed further in this paper. We conclude that there are no physically relevant positive monad bundles on the 101 simple Calabi-Yau  hypersurfaces in smooth toric varieties over and above what has been found for CICYs. 
%

\section{Partial Search: Semi-Positive Monads}
As was mentioned above, unfortunately, the classification programme of positive monads has not given us any new three-generation string models. So, a natural approach to take, in order to find more realistic string models, is to look for bundles under somewhat weaker constraints. The most obvious relaxation is to accept zeros for $\tilde{b}_i^r$ and $\tilde{c}_j^r$, which means that we are searching for semi-positive monads. It is straightforward to see that the classification problem, based on the constraints in Section~\ref{monadsec} but with the positivity condition~\eqref{positivity} replaced by \eqref{spositivity} is no longer closed, in the sense that infinite sets of sums of line bundles $B$ and $C$ compatible with all constraints can be found. The set of associated inequivalent bundles $V$ might still be finite, due to more subtle isomorphisms between monads, but we will not address this somewhat involved problem here. Instead, we ``artificially'' impose the bound $\tilde{S}^r \leq 2$ for all $r$ which leads to a finite search problem for semi-positive monads.

\begin{figure}[t]
\centerline{
(a)
\includegraphics[trim=0mm 0mm 0mm 0mm, clip, width=3.0in]{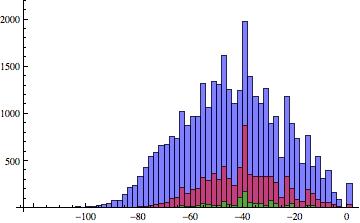}
(b)
\includegraphics[trim=0mm 0mm 0mm 0mm, clip, width=3.0in]{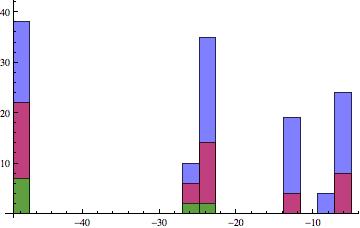}
}
\caption{{\sf The number of semi-positive monads as a function of ${\rm ind}(V)$. Fig.~(a) contains all models, Fig.~(b) only those which satisfy the ``strong'' three-familiy constraint, Eq.~\eqref{constraints}. The three colours blue, red, and green correspond to SU(3), SU(4) and SU(5) models, respectively.}}
\label{f:histo34}
\end{figure}

As before, we impose the following physical constraints on the bundle solutions:
\bea \label{constraints}
\nn 1.&& \text{ind}(V) = 3k\, , k\neq 0 \ , \\
2.&& k ~|~ \chi (X) \ , \\
\nn 3.&& k \text{ belongs to the set, $S(X)$, of possible group orders} \ . 
\eea
in order to filter out candidates for realistic three-generation models. The statistics of semi-positive monads on the 101 simple Calabi-Yau manifolds is summarised in Figure~\ref{f:histo34} and Table \ref{tb4}.

\begin{table}[thb]
{\begin{center}
\begin{tabular}{|c|c|c|c|}\hline 
 &   &  & \\ 
 & $~~~~\text{No Constraints}~~~~$ & $\text{ind}(V)=3k$, $k ~|~ \chi (X)$ & $\text{Constraints Eq.~\eqref{constraints}}$ \\
 &  & &  \\ \hline \hline
$\text{~~SU(3)~~}$ & 35206 & 1902 & 195 \\ \hline
SU(4) & 8066 & 579 & 72 \\ \hline
SU(5) & 1049 & 109 & 13 \\ \hline \hline
Tot. & 44321 & 2590 & 280 \\ \hline
\end{tabular}
\end{center}}
{\caption{\label{tb4} \sf The cumulative number of semi-positive monads on the 101 simple Calabi-Yau manifolds with $\tilde{S}^r \leq 2$. The left column gives the total number of models, the middle column the models satisfying the ``mild'' three-generation constraint ${\rm ind}(V)=3k$, $k\; |\; \chi(X)$ and the right column those which satisfy the ``strong'' three-family constraint, Eq.~\eqref{constraints}.}}
\end{table}

While positive monads existed on only 11 of the 101 base manifolds, semi-positive monads can be found on all spaces. Their number is considerably larger than that of positive monads, as can be seen by comparing Tables~\ref{tb4} with \ref{tb3}. Recall, that in the case of positive monads, there was no model which satisfied the ``strong'' three-generation constraint~\eqref{constraints}. In contrast, we now have $195$ $SU(3)$ models, $72$ $SU(4)$ models and $13$ $SU(5)$ models consistent with this constraint, as Table~\ref{tb4} shows. These models arise on 11 different base manifolds, distributed as shown in Table~\ref{tb5}.
\begin{table}[t!h!b!]
{\begin{center}
\begin{tabular}{|c|c|c|c|c|c|c|c|c|c|c|c|}\hline 
Space No. & $~40_P~$ & $~43~$ & $~56~$ & $~61~$ & $~63~$ & $~69~$ & $~71~$ & $~78~$ & $~105~$ & $~106~$ & $~113~$ \\ \hline \hline
SU(3) & 10 & 3 & 2 &13 & 7 & 32 & 15 &  39 & 6 & 6 & 62\\ \hline
SU(4) & 4 & 1 & 0 & 5 & 2 & 13 & 3 & 19 & 2 & 2 & 21\\ \hline
SU(5) & 1 & 0 & 0 & 1& 0 & 2 & 0 & 5 & 0 & 0 & 4\\ \hline
\end{tabular}
\end{center}}
{\caption{\label{tb5} \sf Number of semi-positive monad solutions with $\tilde{S}^r \leq 2$, which satisfy the ``strong'' three-family constraint~\eqref{constraints}. The subscript $P$ stands for a product of projective spaces.}}
\end{table}

We remark that the bound on $\tilde{S}^r$ was set to 2 merely for practical reasons, in order to keep cpu times in the computer search low. There is no implication that physical models with $\tilde{S}^r>2$ do not exist. In fact, it can be explicitly seen, at least for some base spaces, that this is not the case. For example, as can be seen from Table \ref{tb5}, we have found no three-generation bundles with $\tilde{S}^r\leq 2$ on the space 73, while, for $\tilde{S}^r \leq 3$ there turn out to exist 49, 21 and 6 bundles with structure groups $SU(3)$, $SU(4)$ and $SU(5)$, respectively. Hence, our results do not represent an exhaustive classification of semi-positive three-family models. However, they show that a significant number of promising models do indeed exist. 

Now, let us take a glance at some example solutions. 
We will consider $SU(4)$ semi-positive monads over the space 71 (the seventh column in Table \ref{tb5}), whose set of possible group orders, $S(X)$, turns out to be $\{2,4,8,16\}$.
For this example, $h^{1,1}(\CA) = h^{1,1} (X) = 4$, and therefore, every line bundle is described by a 4-tuple of integers. 
As can be seen in Table \ref{tb5}, there are three $SU(4)$ semi-positive monads over $X$ defined by the exact sequence \eqref{monad}:
\bea
\nn 1: && 0 \to V_1  \to \CO_X (1,0,0,0) \oplus \CO_X (0,1,0,0)^{\oplus 2} \oplus \CO_X (0,0,0,1)^{\oplus2}  \stackrel{f_1}{\to} \CO_X (1,2,0,2) \to 0 \ , \\
\nn 2: && 0 \to V_2  \to \CO_X (1,0,0,0)^{\oplus2} \oplus  \CO_X (0,0,1,0)^{\oplus 2} \oplus \CO(0,0,0,1)^{\oplus 2}  \stackrel{f_2}{\to} \CO_X(1,0,1,1)^{\oplus 2} \to 0 \ , \\
\nn 3: && 0 \to V_3  \to \CO_X (1,0,0,0)^{\oplus2} \oplus  \CO_X (0,1,0,0)^{\oplus 2} \oplus \CO(0,0,0,1)^{\oplus 2}  \stackrel{f_3}{\to} \CO_X(1,1,0,1)^{\oplus 2} \to 0 \ .
\eea
Finally, as the number of $SU(5)$ bundles are reasonably small, we list them exhaustively in Table \ref{tb6}.

\section{Conclusions and Prospects}
In this paper, we have constructed heterotic $E_8 \times E_8$ string models, based on toric Calabi-Yau manifolds and non-trivial vector bundles on them. Specifically, we have restricted our search to a simple class of toric Calabi-Yau manifolds, namely the 101 manifolds which arise as hypersurfaces in smooth toric four-folds and which have 
simplicial K\"ahler cones. Monad bundles with structure group $SU(N)$ (where $N=3,4,5$) have been built over each of these 101 spaces, and a stringent 3-generation constraint (see Eq.~\eqref{constraints}) has been imposed on the resulting models, in order to filter out phenomenologically promising cases.

We have completely classified all positive monads, consistent with heterotic anomaly cancellation, on our 101 base spaces, resulting in a total of 2190 bundles concentrated on just 11 manifolds. From those, only  21 (19 of rank 3, 1 of rank 4, and 1 of rank 5) pass the three-family test, but they all correspond to base spaces which are hypersurfaces in products of projective spaces and have, hence, already been found in the classification of positive monads on CICYs carried out in Ref.~\cite{proj}.
We have then moved on to a partial search of semi-positive monads, which led to a substantially larger list of about 44000 models. Among these,  280 (195 of rank 3, 72 of rank 4, and 13 of rank 5) pass the three-family test.
The 13 semi-positive monads of rank 5 have been listed in Table \ref{tb6}; each of them leads to an $SU(5)$ grand unified theory with three generations. These models, particularly the ones with rank 4 and 5, provide a starting point for the construction of realistic heterotic models on toric Calabi-Yau manifolds with monad bundles.

It is encouraging that even our preliminary scan of the semi-positive bundles has led to a significant number of promising models. It is likely that a more systematic scan, possibly allowing for slightly negative values of the integers $b^r_i$ and $c^r_j$ which specify the monad bundle, will lead to thousands of such models. Such a systematic scan as well as a more detailed analysis of the resulting models will be the subject of future work.

\newpage

\begin{table}[h!!!]
{\begin{center}{
\begin{tabular}{|c|}\hline
Sp No. \\ \hline \hline 
$40_P$ \\[4.2mm] \hline \hline
61 \\[4.2mm] \hline \hline
69 \\[14mm] \hline \hline
78 \\[47mm] \hline \hline
113 \\[57mm]\hline
\end{tabular}
\begin{tabular}{|c|c|c|}\hline 
B & C & ind($V$) \\ \hline \hline 
$ O_X \tiny{\left[\begin{array}{c}1\\[-1mm]0\\[-1mm]0\\[-1mm]0\end{array}\right]} \oplus \tiny{\left[\begin{array}{c}0\\[-1mm]1\\[-1mm]0\\[-1mm]0\end{array}\right]} \oplus  \tiny{\left[\begin{array}{c}0\\[-1mm]0\\[-1mm]1\\[-1mm]0\end{array}\right]}^{\oplus 2} \oplus \tiny{\left[\begin{array}{c}0\\[-1mm]0\\[-1mm]0\\[-1mm]1\end{array}\right]}^{\oplus 2} $  & $O_X \tiny{\left[\begin{array}{c}1\\[-1mm]1\\[-1mm]2\\[-1mm]2\end{array}\right]}$ & -24 \\ \hline

$O_X \tiny{\left[\begin{array}{c}1\\[-1mm]1\\[-1mm]0\\[-1mm]0\end{array}\right]} \oplus \tiny{\left[\begin{array}{c}1\\[-1mm]0\\[-1mm]0\\[-1mm]0\end{array}\right]} \oplus \tiny{\left[\begin{array}{c}0\\[-1mm]1\\[-1mm]0\\[-1mm]0\end{array}\right]} \oplus \tiny{\left[\begin{array}{c}0\\[-1mm]0\\[-1mm]1\\[-1mm]0\end{array}\right]}^{\oplus 2} \oplus \tiny{\left[\begin{array}{c}0\\[-1mm]0\\[-1mm]0\\[-1mm]1\end{array}\right]}^{\oplus 2}$ & $O_X \tiny{\left[\begin{array}{c}1\\[-1mm]1\\[-1mm]1\\[-1mm]1\end{array}\right]}^{\oplus 2}$ & -48 \\ \hline

$ O_X \tiny{\left[\begin{array}{c}1\\[-1mm]0\\[-1mm]0\\[-1mm]0\end{array}\right]} \oplus \tiny{\left[\begin{array}{c}0\\[-1mm]1\\[-1mm]0\\[-1mm]0\end{array}\right]}^{\oplus 2} \oplus \tiny{\left[\begin{array}{c}0\\[-1mm]0\\[-1mm]1\\[-1mm]0\end{array}\right]}^{\oplus 2} \oplus \tiny{\left[\begin{array}{c}0\\[-1mm]0\\[-1mm]0\\[-1mm]1\end{array}\right]}$ & $O_X \tiny{\left[\begin{array}{c}1\\[-1mm]2\\[-1mm]2\\[-1mm]1\end{array}\right]}$  & -48 \\ \hline

$O_X \tiny{\left[\begin{array}{c}1\\[-1mm]0\\[-1mm]0\\[-1mm]0\end{array}\right]}^{\oplus 2} \oplus \tiny{\left[\begin{array}{c}0\\[-1mm]1\\[-1mm]0\\[-1mm]0\end{array}\right]}^{\oplus 2} \oplus \tiny{\left[\begin{array}{c}0\\[-1mm]0\\[-1mm]0\\[-1mm]1\end{array}\right]}^{\oplus 2}$ & $O_X \tiny{\left[\begin{array}{c}2\\[-1mm]2\\[-1mm]0\\[-1mm]2\end{array}\right]}$ & -24 \\ \hline

$O_X \tiny{\left[\begin{array}{c}1\\[-1mm]0\\[-1mm]0\\[-1mm]0\end{array}\right]}^{\oplus 2} \oplus \tiny{\left[\begin{array}{c}0\\[-1mm]1\\[-1mm]0\\[-1mm]0\end{array}\right]} \oplus \tiny{\left[\begin{array}{c}0\\[-1mm]0\\[-1mm]1\\[-1mm]0\end{array}\right]}^{\oplus 2} \oplus \tiny{\left[\begin{array}{c}0\\[-1mm]0\\[-1mm]0\\[-1mm]2\end{array}\right]}$ & $O_X \tiny{\left[\begin{array}{c}2\\[-1mm]1\\[-1mm]2\\[-1mm]2\end{array}\right]}$ & -48 \\ \hline

$O_X \tiny{\left[\begin{array}{c}1\\[-1mm]0\\[-1mm]0\\[-1mm]0\end{array}\right]}^{\oplus 2} \oplus \tiny{\left[\begin{array}{c}0\\[-1mm]1\\[-1mm]0\\[-1mm]0\end{array}\right]} \oplus \tiny{\left[\begin{array}{c}0\\[-1mm]0\\[-1mm]1\\[-1mm]1\end{array}\right]} \oplus \tiny{\left[\begin{array}{c}0\\[-1mm]0\\[-1mm]1\\[-1mm]0\end{array}\right]} \oplus \tiny{\left[\begin{array}{c}0\\[-1mm]0\\[-1mm]0\\[-1mm]1\end{array}\right]}$ & $O_X \tiny{\left[\begin{array}{c}2\\[-1mm]1\\[-1mm]2\\[-1mm]2\end{array}\right]}$ & -48 \\ \hline

$O_X \tiny{\left[\begin{array}{c}1\\[-1mm]0\\[-1mm]0\\[-1mm]1\end{array}\right]} \oplus \tiny{\left[\begin{array}{c}1\\[-1mm]0\\[-1mm]0\\[-1mm]0\end{array}\right]} \oplus \tiny{\left[\begin{array}{c}0\\[-1mm]1\\[-1mm]0\\[-1mm]0\end{array}\right]} \oplus \tiny{\left[\begin{array}{c}0\\[-1mm]0\\[-1mm]1\\[-1mm]0\end{array}\right]}^{\oplus 2} \oplus \tiny{\left[\begin{array}{c}0\\[-1mm]0\\[-1mm]0\\[-1mm]1\end{array}\right]}$ & $O_X \tiny{\left[\begin{array}{c}2\\[-1mm]1\\[-1mm]2\\[-1mm]2\end{array}\right]}$ & -48 \\ \hline

$O_X \tiny{\left[\begin{array}{c}1\\[-1mm]1\\[-1mm]0\\[-1mm]0\end{array}\right]} \oplus \tiny{\left[\begin{array}{c}1\\[-1mm]0\\[-1mm]0\\[-1mm]0\end{array}\right]} \oplus \tiny{\left[\begin{array}{c}0\\[-1mm]0\\[-1mm]1\\[-1mm]0\end{array}\right]}^{\oplus 2} \oplus \tiny{\left[\begin{array}{c}0\\[-1mm]0\\[-1mm]0\\[-1mm]1\end{array}\right]}^{\oplus 2} $ & $O_X \tiny{\left[\begin{array}{c}2\\[-1mm]1\\[-1mm]2\\[-1mm]2\end{array}\right]}$ & -48 \\ \hline

$O_X \tiny{\left[\begin{array}{c}2\\[-1mm]0\\[-1mm]0\\[-1mm]0\end{array}\right]} \oplus \tiny{\left[\begin{array}{c}0\\[-1mm]1\\[-1mm]0\\[-1mm]0\end{array}\right]} \oplus \tiny{\left[\begin{array}{c}0\\[-1mm]0\\[-1mm]1\\[-1mm]0\end{array}\right]}^{\oplus 2} \oplus \tiny{\left[\begin{array}{c}0\\[-1mm]0\\[-1mm]0\\[-1mm]1\end{array}\right]}^{\oplus 2}$ & $O_X \tiny{\left[\begin{array}{c}2\\[-1mm]1\\[-1mm]2\\[-1mm]2\end{array}\right]}$ & -48 \\ \hline

$O_X \tiny{\left[\begin{array}{c}0\\[-1mm]1\\[-1mm]0\\[-1mm]0\\[-1mm]0\\[-1mm]1\end{array}\right]} \oplus \tiny{\left[\begin{array}{c}0\\[-1mm]1\\[-1mm]0\\[-1mm]0\\[-1mm]0\\[-1mm]0\end{array}\right]} \oplus \tiny{\left[\begin{array}{c}0\\[-1mm]0\\[-1mm]0\\[-1mm]1\\[-1mm]0\\[-1mm]0\end{array}\right]}^{\oplus 2} \oplus \tiny{\left[\begin{array}{c}0\\[-1mm]0\\[-1mm]0\\[-1mm]0\\[-1mm]1\\[-1mm]0\end{array}\right]}^{\oplus 2} \oplus \tiny{\left[\begin{array}{c}0\\[-1mm]0\\[-1mm]0\\[-1mm]0\\[-1mm]0\\[-1mm]1\end{array}\right]}$ & $O_X \tiny{\left[\begin{array}{c}0\\[-1mm]1\\[-1mm]0\\[-1mm]1\\[-1mm]1\\[-1mm]1\end{array}\right]}^{\oplus 2}$ & -21 \\ \hline

$O_X \tiny{\left[\begin{array}{c}0\\[-1mm]1\\[-1mm]0\\[-1mm]0\\[-1mm]1\\[-1mm]0\end{array}\right]} \oplus \tiny{\left[\begin{array}{c}0\\[-1mm]1\\[-1mm]0\\[-1mm]0\\[-1mm]0\\[-1mm]0\end{array}\right]} \oplus \tiny{\left[\begin{array}{c}0\\[-1mm]0\\[-1mm]0\\[-1mm]1\\[-1mm]0\\[-1mm]0\end{array}\right]}^{\oplus 2} \oplus \tiny{\left[\begin{array}{c}0\\[-1mm]0\\[-1mm]0\\[-1mm]0\\[-1mm]1\\[-1mm]0\end{array}\right]} \oplus \tiny{\left[\begin{array}{c}0\\[-1mm]0\\[-1mm]0\\[-1mm]0\\[-1mm]0\\[-1mm]1\end{array}\right]}^{\oplus 2} $ & $O_X \tiny{\left[\begin{array}{c}0\\[-1mm]1\\[-1mm]0\\[-1mm]1\\[-1mm]1\\[-1mm]1\end{array}\right]}^{\oplus 2} $ & -21 \\ \hline

$O_X \tiny{\left[\begin{array}{c}0\\[-1mm]1\\[-1mm]0\\[-1mm]0\\[-1mm]0\\[-1mm]1\end{array}\right]}\oplus \tiny{\left[\begin{array}{c}0\\[-1mm]1\\[-1mm]0\\[-1mm]0\\[-1mm]0\\[-1mm]0\end{array}\right]} \oplus \tiny{\left[\begin{array}{c}0\\[-1mm]0\\[-1mm]1\\[-1mm]0\\[-1mm]0\\[-1mm]0\end{array}\right]}^{\oplus 2} \oplus \tiny{\left[\begin{array}{c}0\\[-1mm]0\\[-1mm]0\\[-1mm]0\\[-1mm]1\\[-1mm]0\end{array}\right]}^{\oplus 2} \oplus \tiny{\left[\begin{array}{c}0\\[-1mm]0\\[-1mm]0\\[-1mm]0\\[-1mm]0\\[-1mm]1\end{array}\right]} $ & $O_X \tiny{\left[\begin{array}{c}0\\[-1mm]1\\[-1mm]1\\[-1mm]0\\[-1mm]1\\[-1mm]1\end{array}\right]}^{\oplus 2} $ & -21 \\ \hline

$O_X \tiny{\left[\begin{array}{c}0\\[-1mm]1\\[-1mm]0\\[-1mm]0\\[-1mm]1\\[-1mm]0\end{array}\right]}\oplus \tiny{\left[\begin{array}{c}0\\[-1mm]1\\[-1mm]0\\[-1mm]0\\[-1mm]0\\[-1mm]0\end{array}\right]}\oplus \tiny{\left[\begin{array}{c}0\\[-1mm]0\\[-1mm]1\\[-1mm]0\\[-1mm]0\\[-1mm]0\end{array}\right]}^{\oplus 2} \oplus \tiny{\left[\begin{array}{c}0\\[-1mm]0\\[-1mm]0\\[-1mm]0\\[-1mm]1\\[-1mm]0\end{array}\right]}\oplus \tiny{\left[\begin{array}{c}0\\[-1mm]0\\[-1mm]0\\[-1mm]0\\[-1mm]0\\[-1mm]1\end{array}\right]}^{\oplus 2} $ & $O_X \tiny{\left[\begin{array}{c}0\\[-1mm]1\\[-1mm]1\\[-1mm]0\\[-1mm]1\\[-1mm]1\end{array}\right]}^{\oplus 2}$ & -21 \\ \hline

\end{tabular}
\begin{tabular}{|c|}\hline
Group Order \\ \hline \hline 
2, 4, 8, 16 \\[4.2mm] \hline \hline
2, 4, 8, 16 \\[4.2mm] \hline \hline
2, 4, 8, 16 \\[14mm]\hline \hline
2, 4, 8, 16 \\[47mm]\hline\hline
7\\[57mm]\hline
\end{tabular}

}\end{center}}
{\caption{\label{tb6} \sf Exhaustive list of $SU(5)$ semi-positive monads with $\tilde{S}^r \leq 2$, satisfying the three-generation constraint~\eqref{constraints}; we have marked the simple toric Calabi-Yau spaces in the left-most column, as well as their respective possibilities for orders of freely acting symmetry groups in the right-most. }}
\end{table}

\newpage

\section*{Acknowledgements}
The authors would like to express sincere gratitudes to Maxmilian Kreuzer for the database he shared with us, as well as invaluable correspondences on its interpretation. 
We also acknowledge helpful discussions with Lara Anderson, Philip Candelas, Jeffrey Giansiracusa and James Gray.
Y.-H.~H.~is indebted to the UK STFC for an Advanced Fellowship as well as Merton College, Oxford. 
S.-J.~L.~thanks the Clarendon Fund Bursary and the Overseas Research Scheme for support.

\begin{appendix}
%
\section{Construction of the Manifolds in Toric Geometry} \label{ap:construction}
The three sub--sections in this appendix will constitute a step-wise summary of the construction of our three-folds.
First, we introduce the basic tool kit which will be essential in toric description of varieties, namely, lattices, cones and fans. 
Second, we outline the construction of the toric variety, and finally, we define the Calabi-Yau manifold as a hypersurface in this ambient toric variety.
For a more complete review, the reader can consult \cite{fulton, oda, vafa, agm, cox:review, bouchard, coxnkatz}. 
\subsection{Basic Definitions: Lattices, Cones and Fans}
Let us begin by discussing the spaces on which the toric combinatorial data is defined.
We first introduce a rank $n$ integer lattice $N$ and define its dual lattice $M$ via the natural inner-product $\left\langle \;\cdot \;, \cdot \;\right\rangle : M \times N \rightarrow \IZ$.
Their extensions over $\IR$ are denoted by $N_\IR$ and $M_\IR$, and the same bracket symbol will be used for the extended inner-product. 
We can think of $N$ and $M$ (respectively $N_\IR$ and $M_\IR$) as being isomorphic to $\IZ^n$ (respectively $\IR^n$), and the inner product can be taken as simply the vector dot-product. 
Note that neither the lattices nor their real extensions are directly where the toric variety itself lives; they only furnish as auxiliary spaces. The rank of the lattices, however, is equal to the complex dimension of the toric variety.

Having introduced these objects, we can now define the basic tool-kit. A set $\sigma \subset N_{\IR}$ is a \textit{strongly convex rational polyhedral cone} if 
\be 
\sigma = \left\{ \sum\limits^{k}_{i=1}  a_i \textbf v_i  ~|~ a_i \in \IR_{\geq 0} \right\} 
\ee 
for a finite set of vectors $\textbf v_1, \cdots ,\textbf v_k$ $\in N$ and $\sigma \cap (-\sigma) = \left\{ \bold{0} \right\}$.
For simplicity, $\sigma$ is often called a \textit{cone}.  
Every cone $\sigma \subset N_{\IR}$ has its \textit{dual cone} $\check{\sigma} \subset M_{\IR}$ defined as
\be 
\check{\sigma} = \left\{\bold m \in M_{\IR} ~|~ \left\langle \bold m, \bold v\right\rangle \geq 0 \;\;\forall\, \bold v \in \sigma \right\} \ .
\ee 
A set $\tau \subset \sigma$ is called a \textit{face} of the cone $\sigma$ if it is spanned over $\IR_{\geq 0}$ by a subset of generators of $\sigma$ and lies on the boundary of $\sigma$. 
A \textit{fan} is then defined as a collection $\Sigma$ of cones in $N_{\IR}$ such that each face of a cone in $\Sigma$ is also a cone in $\Sigma$ and the intersection of two cones in $\Sigma$ is a face of each. 
The collection of $d$-dimensional cones in $\Sigma$ is denoted by $\Sigma^{(d)} \text{~for~} 0 \leq d \leq n$. 
We also denote the union of all the cones in $\Sigma$ by $|\Sigma|$ and call it the \textit{support} of $\Sigma$. 

\begin{figure}[h!]
\centerline{
\includegraphics[trim=0mm 0mm 0mm 0mm, clip, width=6.0in]{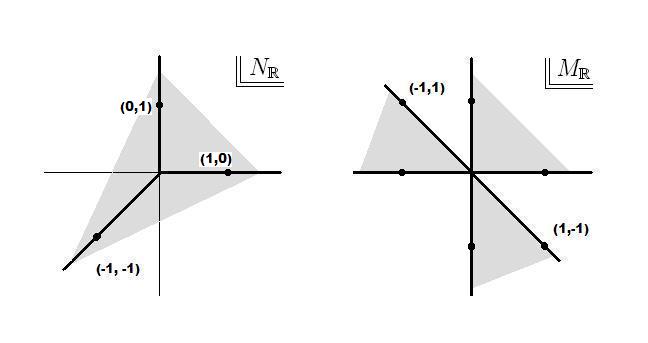}}
\caption{\sf The fan for $\IP^2$ (left) and the 2-dimensional dual cones (right).}
\label{f:fan}
\end{figure}
As an example, Figure \ref{f:fan} depicts a fan in $N_\IR = \IR^2$, shown at the left.
This fan consists of one 0-dimensional cone, namely, the origin $(0,0)$,  three 1-dimensional cones, namely the three rays generated respectively by $\bold v_1 = (1,0), ~\bold v_2 = (0,1), ~\bold v_3 = (-1,-1)$, as well as three 2-dimensional cones (shaded), generated respectively by the neighbouring pairs: $\{\bold v_1, \bold v_2\}$, $\{\bold v_2, \bold v_3\}$, and $\{\bold v_3, \bold v_1\}$.
The three 2-dimensional dual cones are depicted on the right. 

\subsection{Construction of Toric Varieties}\label{ap:toric}
There are several equivalent ways how we construct toric varieties from their toric data, that is from their associated fans. Amongst them is the algebro-geometric construction, where each affine patch of the variety is explicitly realised as the maximal spectrum of some ring.
One of the basic ideas underlying this local construction is that there is an one-to-one correspondence between the cones $\sigma \in \Sigma$ and the orbits of the torus action $T$ on the toric variety $\CA$. 
It turns out that the correspondence is dimension-reversing. 
To be precise, 
\be
{\rm dim} (\sigma) + {\rm dim} ({\rm orb} (\sigma) ) = n.
\label{corresp}
\ee
In particular, the top-dimensional cones correspond to the fixed points of the $T$-action and the 1-dimensional cones to the $T$-invariant divisors. 
We denote such divisors by $D_\rho$ where $\rho \in \Sigma^{(1)}$ are the edges in the fan. 

In this paper, however, we are more interested in the global construction.
Let us first recall the ordinary construction of $\IP^n$. 
One considers $\IP^n$ as the quotient of $\IC^{n+1} - \{\bold{0}\}$ by the multiplicative group $\IC^*$. 
Each point in $\IP^n$ is labelled by its homogeneous coordinates $(x_1, \cdots , x_{n+1})$, which we identify with $\lambda \cdot (x_1, \cdots , x_{n+1})$ for all $\lambda \in \IC^{*}$. 
This can be easily generalised to the case of arbitrary toric varieties. 

With each edge $\rho \in \Sigma^{(1)}$ of the fan $\Sigma$, we associate a \textit{homogeneous coordinate} $x_\rho$. 
So there are $k$ homogeneous coordinates $(x_1, \cdots, x_k)$ on $\mathbb{C}^k$, where $k=|\Sigma^{(1)}|$.
Just as for ordinary projective spaces, the next task is to identify certain measure zero subsets of $\mathbb{C}^k$ which should be removed. Let $S$ be a subset of $\Sigma^{(1)}$ that does not span a cone of $\Sigma$ and let $Z(S) \subset \IC^k$ be the linear subspace defined by setting $x_{\rho} = 0\;\; \forall\, \rho \in S$. 
Now let $Z(\Sigma) \subset \IC^k$ be the union of all such subspaces $V(S)$. 
Then the toric variety is constructed as a quotient of $\IC^n - Z(\Sigma)$ by some group G. 
We refer to \cite{vafa} for a detailed description of how $G$ is constructed. Here we rather content ourselves with a partial answer which is valid for the smooth toric varieties which are the primary interest of the present paper. 

For such cases, $G$ is isomorphic to $(\IC^*)^{k-n}$ and the $G$ quotient is implemented by the following equivalence relations
\be
(x_1, \cdots, x_k) \sim (\lambda_r^{\beta^r_{~1}} x_1, \cdots, \lambda_r^{\beta^r_{~k}} x_k) \ ,
\ee
with $\lambda_r \in \IC^*$. The coefficients $\beta^r_{~\rho}$ are defined by the linear relations $\sum \limits_{\rho=1}^{k} \beta^r_{~\rho} \bold v_\rho =0$ which amount to $n$ independent conditions. Hence, $\beta^r_{~\rho}$ form an $(k-n) \times k$ matrix which is often referred to as a {\bf charge matrix} \cite{vafa}. Choosing all its entries to be integers and requiring that $\rm{g.c.d.} (\beta^r_{~1}, \cdots, \beta^r_{~k}) = 1$ it is uniquely defined (up to lattice isomorphisms). It is easy to see that $G$ preserves $\IC^k - Z(\Sigma)$ and hence, we can take the quotient \be \CA = (\IC^k-Z(\Sigma))/G \ , \ee to construct the toric variety. 

\subsection{Construction of Calabi-Yau Hypersurfaces}
In this sub--section, we briefly describe how to construct our desired Calabi-Yau three--fold $X$ as a hypersurface of  a 4-dimensional ambient toric variety $\CA$.

Not every toric $n$-fold contains a Calabi-Yau hypersurface. To formulate what exactly the condition on the fan is, we first need to introduce an $n$-dimensional {\bf polytope} $\Delta \subset M_\IR$.
By a polytope, we mean that $\Delta$ is the convex hull of a certain set, which one can take to be the set of vertices of $\Delta$. This is called the vertex representation, for the obvious reason. 
As an equivalent definition, a polytope can also be defined as the intersection of a finite number of half-spaces, which can be chosen as the collection of facet-defining half-spaces. 
We only consider a polytopes containing the origin and hence, can subsequently define its {\bf{dual polytope}} $\Delta^\circ \subset N_\IR$ as 
\be
\Delta^\circ = \{ \bold{v} \in N_{\IR} ~|~ \left\langle \bold{m},\bold{v} \right\rangle \geq -1\;\;{\forall}\, \bold{m} \in \Delta \} \ .
\ee
The polytope $\Delta$ is called {\bf{reflexive}} if all the vertices of $\Delta$ as well as $\Delta^\circ$ are lattice points. 
Note that the dual polytope $\Delta^\circ \subset N_\IR$ also contains the origin as its interior point. 
We can then define a fan $\Sigma$ in $N_\IR$ which consists of the cones over the faces of $\Delta^\circ$ with their apexes at the origin. 
This fan $\Sigma$ is called the {\bf normal fan} of the polytope $\Delta$, and we have the following statement:
\textit{the normal fan $\Sigma$ in $N_\IR$ of a reflexive polytope $\Delta \subset M_\IR$ defines a toric n-fold as well as a Calabi-Yau (n-1)-fold embedded therein.}

More precisely, the normal fan tells us about the defining equation of the Calabi-Yau hypersurface as follows.
To each lattice point $\bold m$ of a reflexive polytope $\Delta \subset M_{\IR}$ we assign a monomial 
\be
\label{monomial} 
\bold{x}^{[\bold m]} = \prod\limits_{\rho = 1} ^{k} {x_{\rho} ^ {\left\langle \bold m, \bold v_\rho \right\rangle + 1}}, 
\ee 
where $x_{\rho=1,\cdots, k}$ are the homogeneous coordinates of the toric variety $\CA$ associated to the polytope $\Delta$. These homogeneous coordinates correspond to the $k$ edge vectors $\bold v_{\rho=1, \cdots, k}$ of the normal fan $\Sigma$ of $\Delta$. Now, it turns out that a linear combination of all the monomials corresponding to the lattice points $\bold{m} \in \Delta$ is a homogenous polynomial and hence, its zero locus can define a hypersurface $X$ to $\CA$. 
What is more, the hypersurface $X$ indeed satisfies the Calabi-Yau condition. 
It is straightforward to see that this defining polynomial is a section of the line bundle $\mathcal{O}_{\CA} (\sum\limits_{\rho=1}^{k} D_{\rho} )$, the anticanonical bundle of the ambient space $\cal A$. 
In other words, the normal bundle of $X$ is 
\be \label{nb} \mathcal {N} = \mathcal O _{\CA} (\sum\limits_{\rho=1}^{k} D_\rho) \ . \ee 

\begin{figure}[h!]
\centerline{
\includegraphics[trim=0mm 0mm 0mm 0mm, clip, width=6.0in]{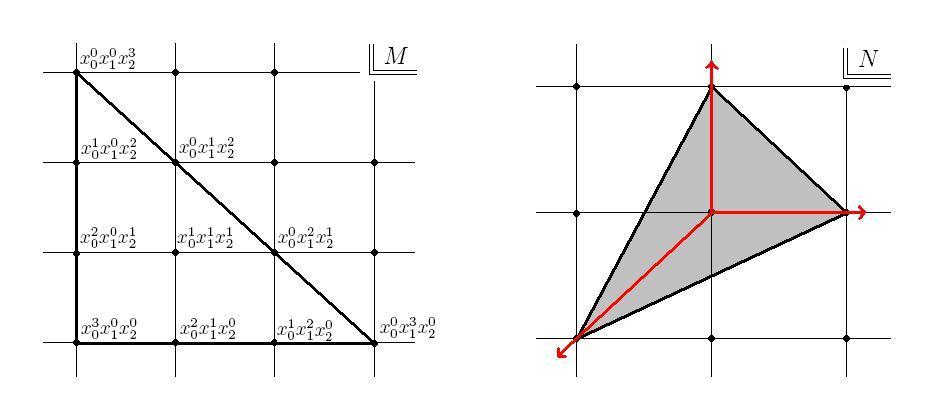}}
\caption{\sf A polytope $\Delta \subset M_\IR$(left) and its dual polytope $\Delta^\circ \subset N_\IR$(right).}
\label{f:polytope1}
\end{figure}
Figure \ref{f:polytope1} is a 2-dimensional example depicting a reflexive polytope $\Delta \subset M_\IR$ and the dual polytope $\Delta^\circ \subset N_\IR$. 
Note that the normal fan $\Sigma$ of $\Delta$, which is the collection of the cones over the faces of $\Delta^\circ$, reproduces the fan for $\IP^2$ in Figure \ref{f:fan}. 
Note also that the polytope $\Delta \subset M_\IR$ in the figure gives us all the monomials of degree 3 and hence, defines the toric variety $\IP^2$ as well as the family of cubic Calabi-Yau hypersurfaces. 
Of course, the lattice of our main concern is of rank 4, rather than of rank 2 as in this simple example. 
It turns out that there are $473,800,776$ 4-dimensional reflexive polytopes \cite{reflexive} and hence, that many Calabi-Yau 3-folds arise this way.

\section{Relevant Properties of the Manifolds} \label{ap:properties}
Various properties of the ambient toric varieties and their associated Calabi-Yau hypersurfaces can be easily read off from the toric data. 
Here, we summarise the ones relevant to our string models. 

Given a toric variety constructed by its fan, a natural question to ask is how we describe line bundles thereon; this will be key to our discussion of monads.
We have a simple answer to this question for a smooth, compact, toric variety.
The Picard group $\text{Pic}(\CA)$, which parametrises the space of line-bundles on $\CA$, is determined by the following short exact sequence 
\be 
\label{se}
0 \rightarrow M \stackrel{\alpha}{\rightarrow} \bigoplus\limits_{\rho=1}^{k} \IZ D_\rho \stackrel{\beta}{\rightarrow} \text{Pic}(\mathcal A)  \rightarrow 0
\ee
where $k= |\Sigma^{(1)}|$ is the number of edges in $\Sigma$ and $n=\text{dim}_{\IC} \CA$, we recall, is the complex dimension of $\CA$. 
The first map $\alpha$ maps $\bold m$ to $\sum\limits_{\rho=1}^{k} {\left\langle \bold m , \bold {v_\rho} \right\rangle D_\rho }$ and therefore, 
\be
\nn
\text{Ker} (\beta) = \text{Im} (\alpha) = \left\{(\left\langle \bold m , \bold v_1 \right\rangle , \cdots ,\left\langle \bold m , \bold v_k \right\rangle) ~|~ \bold m \in M \right\}\; .
\ee 
This expression for $\text{Ker}(\beta)$ together with the exactness of the sequence~\eqref{se} fixes the linear map $\beta$ up to lattice isomorphisms. In fact, the $(k-n) \times k$ matrix representing the $\beta$-map, is precisely the charge matrix $\left[\beta^r_{~\rho} \right]$ defined in \ref{ap:toric}. Since the dual lattice $M$ is isomorphic to $\IZ^n$, the short exact sequence \eqref{se} implies \be 
\text{Pic} (\CA) \simeq \IZ^{k-n} \ .
\ee 
So every line bundle is determined by a $(k-n)$-tuple of integers, and we can denote it by $\cO_{\CA} (\bold a)$ for $\bold a \in \IZ^{k-n}$.  A basis of $(1,1)$-forms for $H^2(\CA, \IZ)$ can then be defined by setting $J_r \equiv c_1 ( \mathcal O_{\CA} ( \bold e_r ) )$ for $r=1, \cdots , k-n$, where ${\bf e}_r$ are the standard unit normal vectors in $\mathbb{Z}^{k-n}$. With this definition the first Chern class of line-bundles can be written as
\begin{equation}
c_1 (\CO_{\CA} (\bold{a}) ) = a^r J_r \ , 
\end{equation}
where the sum over the index $r $ is implicit.

The non-trivial Hodge numbers of the smooth Calabi-Yau 3-fold $X$ are given by the formulas~\cite{batyrev}
\bea
\label {hodge11}
h^{1,1} (X) &=& \textit{l}(\Delta^{\circ}) - 5 - \sum\limits_{\text{codim} \check{\Theta} = 1} \textit{l}^\star ( \check{\Theta}) + \sum\limits_{\text{codim} \check{\Theta} = 2} \textit{l}^\star (\check{\Theta} ) \textit{l}^\star (\Theta) \ , \\
\label {hodge21}
h^{2,1} (X) &=& \textit{l}(\Delta) - 5 - \sum\limits_{\text{codim} \Theta = 1} \textit{l}^\star (\Theta) + \sum\limits_{\text{codim} \Theta = 2} \textit{l}^\star (\Theta ) \textit{l}^\star (\check{\Theta}) \ .
\eea
Here, $\textit{l} (\Theta)$ denotes the number of lattice points in $\Theta$, and $\textit{l}^{\star} (\Theta)$ the number of lattice points in the interior of $\Theta$. 
The summations run over the faces $\Theta$ and $\check{\Theta}$ of the polytopes $\Delta$ and $\Delta^\circ$, respectively. As was mentioned in the main text of this paper, it turns out that all the pairs of $\CA$ and $X$ within our database satisfy $\rm{dim}(\text{Pic}(\CA)) = h^{1,1}(X)=h^{1,1}({\cal A})$. For simplicity, we will denote this number by $h^{1,1}$. 

Another important task is to compute the intersection numbers of both the ambient space $\CA$ and the Calabi-Yau hypersurface $X$. 
We first work out the intersection numbers of $\CA$
\beq 
d_{rstu}= \int_{\cal A} J_r \wedge J_s \wedge J_t \wedge J_u \ ,
\eeq
where $r,s,t,u = 1, \cdots , h^{1,1}$.
The basic idea is to take four edge vectors of the fan and check whether they span a four-cone or not.
A linear equation on $d_{rstu}$ arises from the choice of the four distinct edges $\rho_1, \rho_2, \rho_3, \rho_4$ as follows: 
\be \label{inter}
d_{rstu} \beta^r_{~\rho_1} \beta^s_{~\rho_2} \beta^t_{~\rho_3} \beta^u_{~\rho_4} = \left\{ 
\begin{array}{l l}
1 & \quad \mbox{if \{$\bold v_{\rho_1}, \bold v_{\rho_2}, \bold v_{\rho_3}, \bold v_{\rho_4}$\} spans a 4-cone \ ,}\\
0 & \quad \mbox{otherwise \ ,}\\ 
\end{array} \right. 
\ee
where the summations over $r,s,t,u$ are implicit. Even if a vector appears multiple times in the set $\{\bold v_{\rho_1}, \bold v_{\rho_2}, \bold v_{\rho_3}, \bold v_{\rho_4}$\} Eq.~\eqref{inter} still holds provided the set does not span a cone. By making different choices for the set of vectors we can obtain a set of simultaneous equations which uniquely determine the intersection numbers $d_{rstu}$. It is then straightforward to calculate the intersection numbers $d_{rst}$ of the (favourable) Calabi-Yau hypersurface $X$ by
\bea \label{interx}
\nn d_{rst} &=& \int_X J_r \wedge J_s \wedge J_t \\
 &=& \int_{\mathcal A} J_r \wedge J_s \wedge J_t  \wedge c_1 (\mathcal N) \\
\nn &=& n^u d_{rstu}
\eea 
where $c_1(\mathcal N) := n^u J_u$. Note that, by abuse of notation, we denote the (1,1) forms on $\CA$ and their pull-backs to $X$ by the same symbol $J_r$.

We now move on to Chern classes. 
The total Chern class of $\CA$ is given by 
\be
\label{chern1}
c(\CA)= \prod\limits_{\rho=1}^{k} [1+c_1(\CO_\CA ( D_\rho))]
\ee
where $\CO_\CA ( D_\rho)$ is the line-bundle associated to the toric divisor $D_\rho$. 
On the other hand, the correspondence between divisors and line-bundles can be inferred from the $\beta$-map which appears in \eqref{se}. The expression \eqref{chern1} for the Chern class then simplifies to
\be
\label{chern2}
c(\CA) = \prod\limits_{\rho=1}^{k} \left[ 1+ \beta^{r}_{~\rho} J_r \right],
\ee
where $J_r \equiv c_1 (\CO_\CA ( \bold{e}_r))$ for $r=1, \cdots, k-n$. For instance, for the first two non-trivial terms in \eqref{chern2} one reads off
\bea
\nn c_1(\CA) &=& \sum\limits_{1\leq \rho \leq k}^{} {\beta^r_{~\rho} J_r} \ , \\
\nn c_2(\CA) &=& \sum\limits_{1\leq \rho < \sigma \leq k}^{} {\beta^r_{~\rho} \beta^s_{~\sigma} J_r J_s} \ .
\eea

On the other hand, we have the following standard short exact sequence
\begin{equation}
0 \rightarrow TX \rightarrow T\mathcal{A}|_X \rightarrow \mathcal{N} \rightarrow 0 \ ,
\end{equation}
which relates the tangent bundles $TX$ of our Calabi-Yau threefold $X$, the restriction $T\CA|_X$ of the tangent bundle $T\CA$ of ${\cal A}$ to $X$ and the normal bundle $\CN$ of $X$ in $\CA$. 
The above sequence implies that the Chern classes of these three bundles are related by
\be \label{chernx}
c(\mathcal A) = c(X) \wedge c(\mathcal N) \ .
\ee
This relation can also be understood in terms of the adjunction formula.
Combining the result with Eqs.~\eqref{nb} and \eqref{chern2}, it is straightforward to calculate $c(X)$, and in particular, $c_2(X)$, which, in fact, turns out to be equal to $c_2(\cal A)$.  

In the rest of this section, we study K\"ahler and Mori cones.
As a preparation, we cite the following theorem
\begin{theorem}
The toric variety of a fan $\Sigma$ in $N_\IR$ is projective if and only if $\Sigma$ is the normal fan of an $n$-dimensional lattice polytope $\Delta \subset M_\IR$.
\end{theorem}
which assures us that ${\cal A}$ always admits K\"ahler structures.

In order to determine the K\"{a}hler cone of $\CA$, we first associate to each cohomology class $\bold a = \sum\limits_{\rho=1}^{k}{a^\rho \left[ D_\rho\right]} \in H^{1,1}(\CA, \IR)$, a {\bf support function} $\psi : |\Sigma| \rightarrow \IR$ defined as follows.
For every maximal cone $\sigma \in \Sigma$, there is a unique $\bold m_\sigma \in M_\IR$ such that
\be 
\label{support_def} \left\langle \bold m_\sigma , \bold v_\rho \right\rangle = - a^\rho \text{ if } \bold v_\rho \subset \sigma \ ,
\ee 
and extending this linearly over the cone $\sigma$ we can define a linear function on $\sigma \subset \Sigma$. 
Now, with this as a local definition, we construct the support function $\psi$ on the whole support $|\Sigma|$, which can be thought of as the union of all maximal cones of $\Sigma$. More precisely, we define the $\Sigma$-piecewise linear function $\psi : |\Sigma| \rightarrow \IR$ so that 
\be 
\psi (\bold v) = \left\langle \bold m_\sigma, \bold v \right\rangle \ ,
\ee
where $\sigma$ is a maximal cone containing $\bold v$.
Note that $\psi(\bold v)$ has a well-defined value even when $\bold v$ is contained in more than one maximal cones, due to Eq.~\eqref{support_def} and to the linearity over each $\sigma$.
We call $\psi$ the support function of the class $\bold a$. 
The cohomology class $\bold a$ is said to be {\bf convex} if its support function $\psi$ is a convex function in the usual sense.\footnote{A real-valued function $f$ defined on a convex subset $C \subset \IR^n$ is called \emph{convex} if, for any two points $\bold x$ and $\bold y$ in its domain $C$ and any $t$ in $[0,1]$, we have $f(t \bold x + (1-t) \bold y ) \leq t f(\bold x) + (1-t) f(\bold y)$.}
Convex classes form a cone denoted by $\rm{cpl} \subset H^{1,1}(\CA, \IR)$. 
Now, the following theorem determines the K\"{a}hler cone of $\CA$: 
\begin{theorem} 
\label{cpl}
If $\cal A$ is a simplicial projective toric variety\footnote{A toric variety is simplicial if each cone in its fan is simplicial, \textit{i.e.}, if the generators of each cone are linearly independent.}, then $\emph{cpl}(\Sigma) \subset H^{1,1} (\mathcal A, \IR)$ is a strongly convex polyhedral cone with nonempty interior in $H^{1,1} (\mathcal A, \IR)$. 
Furthermore, the interior of this cone is precisely the K\"{a}hler cone of $\cal A$.
\end{theorem}
Support functions $\psi$ corresponding to K\"{a}hler classes are then \emph{strictly} convex. Thus, the theorem below provides the practical prescription for the K\"ahler cone: 
\begin{theorem} 
\label{kahler}
If $\cal A$ is a simplicial projective toric variety, then the support function $\psi$ of $\sum\limits_{\rho=1}^{k} {a^\rho [D_\rho]}$ is strictly convex if and only if for every primitive collection\footnote{A primitive collection of a fan $\Sigma$ is a subset $\mathcal{P} \subset \Sigma^{(1)}$ s.t. $\mathcal{P}$ itself is not the set of generators of a cone in $\Sigma$ while every proper subset of $\mathcal{P}$ is.} $\mathcal{P} = \{\bold v_1, ... , \bold v_l\}$, we have
\be \label{strict} \psi(\bold v_1 + ... + \bold v_l) > \psi(\bold v_1) + ... + \psi (\bold v_l) \ . \ee 
\end{theorem}
For each primitive collection $\mathcal{P}$, Eq.~\eqref{strict} gives a linear homogeneous inequality for $a^\rho$, which then leads to the corresponding inequality for the K\"ahler moduli $t^r$. Here, we make use of the map $\beta$, which relates $a^\rho$ linearly to $t^r$. Now we scan over all the primitive collections of the fan $\Sigma$ and choose a maximal set of the independent inequalities. 
This set forms a system of linear homogeneous inequalities on $t^r$ which can be written as
\begin{equation}
 K^{\bar{r}}_{~r} t^r \geq 0 \mbox{ for } \bar{r}=1, \cdots , m\; .
\end{equation} 
with an $m \times h^{1,1}$ matrix $K=\left[ K^{\bar{r}}_{~r}\right]$, where $m$ is the cardinality of the maximal set.

For a favourable Calabi-Yau hypersurface $X$, every closed $(1,1)$-form in $X$ can be thought of as the pull-back of a $(1,1)$-form in $\cal A$. 
Hence, the K\"{a}hler cone of $X$ must contain that of $\cal A$ (note the reverse inclusion). 
Although we do not have a complete understanding of the K\"{a}hler cone of $X$, it is plausible to conjecture that for smooth toric ambient spaces the K\"{a}hler cone of $X$ is equal to that of $\cal A$. 
We will work under this assumption when we need the precise details of the K\"{a}hler cone of $X$.

The set of effective curves in a K\"ahler manifold generates a cone; these live in $H_2(\CA, \IZ)$ and form a cone which is dual to the K\"ahler cone. This cone is called the {\bf Mori cone} of $\CA$.
Of course, once the K\"{a}hler cone is known, the Mori cone can be obtained as its dual.
On the other hand, the toric data provides an alternative way of calculating the Mori cone \cite{reid}, and this can serve us as a consistency check for our K\"{a}hler cone calculations. Indeed, we have confirmed that each of the edge vectors of the Mori cone corresponds to a facet of the K\"ahler cone.

Now, the Mori cone of $X$ should be contained in that of $\cal A$ due to the duality of Mori and K\"{a}hler cones. 
We assume that the two Mori cones are the same for our collection of smooth spaces. 

\section{The Database and an Illustrative Example} \label{ap:database}
Table \ref{tb7} lists the complete database of the 124 smooth toric 4-folds which contain the Calabi-Yau 3-folds; the two hodge numbers of the Calabi-Yaus are denoted below the space number as $(h^{1,1}, h^{2,1})$. 
The toric data is expressed in terms of the 4-dimensional reflexive polytopes $\Delta^\circ \subset N_\IR$. 
For reference, we separately tabulate in Table \ref{tb8} those ambient spaces which are products of del Pezzo surfaces and projective spaces.

As outlined previously, the polytope information is sufficient in order to determine all the relevant differential-geometric properties of the ambient and Calabi-Yau spaces.
Let us illustrate this by an explicit example. 
The two simplest spaces in Table~\ref{tb7}, with labels $1_P$ and $2$ correspond respectively to the quintic hypersurface in $\IP^4$ and the bidegree-(3,3) hypersurface in $\IP^2 \times \IP^2$.  Hence, we will work with the next simplest and non-trivial space with label $3$. 

\paragraph{Fan:}
The toric data in Table \ref{tb7} shows the lattice vertices of $\Delta^\circ \subset N_\IR$. 
Because the normal fan $\Sigma$ in $N_\IR$ consists of the cones over the faces of $\Delta^\circ$, the lattice vertices of $\Delta^\circ$ correspond precisely to the edge vectors of $\Sigma$. Hence, the set of one-cones can be directly read off from Table \ref{tb7}:
\be \nn 
\bold v_1 = \bold e_1; ~\bold v_2 = \bold e_2; ~\bold v_3 = \bold e_3; ~\bold v_4 = \bold e_4; 
~\bold v_5 = (-1,~0,~0,~0); ~\bold v_6 = (~1,-1,-1,-1) \ .
\ee
Here, $\bold e_1, \cdots,  \bold e_4$ are the standard unit vectors.
From these, one can also compute the higher dimensional cones, and this process has already been automated in the computer programme PALP \cite{palp}. Figure \ref{f:palp} shows the PALP input and output screen for our example, and it lists all the cones in the normal fan $\Sigma$.
\begin{figure}[h!]
{\scriptsize
\begin{verbatim}
Degrees and weights  `d1 w11 w12 ... d2 w21 w22 ...'
  or `#lines #colums' (= `PolyDim #Points' or `#Points PolyDim'):
4 6 
Type the 24 coordinates as dim=4 lines with #pts=6 colums:
1 0 0 0 -1 1 
0 1 0 0 0 -1
0 0 1 0 0 -1
0 0 0 1 0 -1
M:7 6 N:111 8 H:90,2 [176]
Incidences as binary numbers [F-vector=(6 14 16 8)]:
v[d][i]: sum_j Incidence(i'th dim-d-face, j-th vertex) x 2^j
v[0]: 100000 001000 010000 000010 000100 000001 
v[1]: 101000 110000 011000 100010 001010 010010 100100 001100 010100 100001 001001 000011 000101 000110 
v[2]: 111000 101010 110010 011010 101100 110100 011100 101001 100011 001011 100101 001101 100110 010110 000111 001110 
v[3]: 111010 111100 101011 101101 110110 100111 011110 001111 
\end{verbatim}
}
\caption{\sf The in/out-put screen in PALP \cite{palp}. The first input 4 and 6 denote the lattice rank and the number of the vertices in $\Delta^\circ$, respectively, and the second input is the list of those vertices, $\rho$-th column being $\bold v_\rho$ for $\rho=1, \cdots, 6$. The output includes two Hodge numbers and Euler character of $X$, which are denoted by H in the middle, as well as the incidence information of the normal fan $\Sigma$. The latter is expressed in binary notation: for instance, the first entry in the last row, 111010, represents a four-cone generated by the four edge vectors $\bold v_2, \bold v_4, \bold v_5$ and $\bold v_6$.}
\label{f:palp}
\end{figure}

PALP also has the routine for calculating the hodge numbers of $X$ and the result is, as shown in Figure \ref{f:palp},
\be \nn
h^{1,1} = 2; ~ h^{1,2} = 90 \ .
\ee
We could as well work out these numbers by hands, using Eqs.~\eqref{hodge11} and \eqref{hodge21}. 

\paragraph{Charge Matrix:}
The six edge vectors $\bold v_1, \cdots, \bold v_6$ have two linear relations 
\bea
\nn \bold v_1 + \bold v_5 &=&0 \ , \\
\nn -\bold v_1 + \bold v_2+ \bold v_3+ \bold v_4 + \bold v_6 &=&0 \ ,
\eea
and thus, we have the following charge matrix:
\be \nn
\beta = \left( 
\begin{array}{cccccc} 
1 & 0 & 0 & 0 & 1 & 0 \\
-1 & 1 & 1 & 1 & 0 & 1 
\end{array}
\right) \ .
\ee
So, the divisor-linebundle correspondence follows from the short exact sequence \eqref{se}, which tells us that $\text{Pic} (\CA) \simeq \IZ^2$ and that the divisor $D=a^\rho D_\rho$ corresponds to the line-bundle \be \beta (D) = \beta \cdot \bold a = \left( \begin{array}{c} \beta^1_{~\rho} a^\rho \\ \beta^2_{~\rho} a^\rho \end{array} \right) \ , \nn \ee
where sums over $\rho$ are implicit.

\paragraph{Normal Bundle:}
The normal bundle $\CN$ of the Calabi-Yau hypersurface is, by Eq.~\eqref{nb}, the line-bundle corresponding to the divisor $D_N = D_1 + \cdots + D_6$, which gets mapped by $\beta$ to the 2-tuple:
\be
\nn
\beta (D_N) = \left( \begin{array}{c} 2 \\ 3 \end{array} \right) \ .
\ee
Thus, the normal bundle is represented as 
\be \nn \CN = {\cal O}_{\cal A}(2,3) \ , \ee
and hence, bi-degree $(2,3)$ homogeneous equations define the family of our Calabi-Yau hypersurfaces in this toric variety. For instance, the monomial corresponding to the origin $\bold{0} \in \Delta$ is, by \eqref{monomial}, 
\be \nn
\bold x^{[\bold 0]} = x_1 x_2 x_3 x_4 x_5 x_6 \ ,
\ee
whose bi-degree $(a, b)$ is obtained as
\bea 
\nn a &=& 1+ 0 + 0+ 0 + 1 + 0 = 2 \ , \\
\nn b &=& -1 + 1 + 1 + 1 + 0 + 1 = 3  \ .
\eea
Note that the degrees are added up weighted by the entries of the charge matrix $\beta$.
One can check that every lattice point in $\Delta$ corresponds to a monomial of the same bi-degree. 

\paragraph{Intersection Numbers:}
The intersection numbers $d_{rstu}$ in $\CA$ have 5 degrees of freedom, namely, $d_{1111}, d_{1112}, d_{1122}, d_{1222}$ and $d_{2222}$. 
Thus, we have to make at least five choices of four edge vectors in the fan, in order to obtain five linear equations of the form \eqref{inter}. Many of these are redundant and five independent choices are:
\bea 
\nn\{ \bold v_1, \bold v_2, \bold v_3, \bold v_5 \} &\rightarrow & 010111 ~\rightarrow ~ 0 = d_{1122}-d_{1222} \\
\nn\{ \bold v_2, \bold v_3, \bold v_4, \bold v_5 \} &\rightarrow & 011110 ~\rightarrow ~ 1 = d_{1222} \\
\nn\{ \bold v_2, \bold v_3, \bold v_4, \bold v_6 \} &\rightarrow & 101110 ~\rightarrow ~ 0 = d_{2222} \\
\nn\{ \bold v_1, \bold v_5, \bold v_5, \bold v_5 \} &\rightarrow & 010001 ~\rightarrow ~ 0 = d_{1111} - d_{1112} \\
\nn\{ \bold v_1, \bold v_1, \bold v_5, \bold v_5 \} &\rightarrow & 010001 ~\rightarrow ~ 0 = d_{1111} - 2 d_{1112} + d_{1122} \ .
\eea
Note that the middle column is written in binary notation so that we can check with the incidence information shown in Figure \ref{f:palp}, and that Eq.~\eqref{inter} has been used in the last step.
The solution to the above set of simultaneous equations is
\be \nn
d_{1111}= 1;~d_{1112}= 1;~d_{1122}=1 ;~d_{1222}=1 ;~d_{2222}=0 \ .
\ee
Now, the intersection numbers $d_{rst}$ in $X$ are, from Eq.~\eqref{interx}, (with $n^1=2$ and $n^2=3$) 
\be \nn
d_{111}=5;~d_{112}=5;~d_{122}=5;~d_{222}=2 \ .
\ee

\paragraph{Chern Class:}
The total Chern class of $\CA$ is directly given by \eqref{chern2}
\bea
\nn c(\CA)&=&(1+ J_1 - J_2) ( 1 + J_2)^4 (1+J_1) \\
\nn &=&1+(2J_1 + 3J_2) + (J_1^2 + 7 J_1J_2 + 2J_2^2) + (4J_1^2 J_2 + 8 J_1 J_2^2 -2 J_2^3 ) + (6J_1^2J_2^2 + 2J_1 J_2^3 - 3J_2^4) \ ,
\eea
from which all the Chern classes can be read off. 
The relation \eqref{chernx} between $c(X), c(\CA)$ and $c(\CN)$ can then be used to compute the total Chern class of $X$:
\bea
\nn c(X) = \frac{c(\CA)}{1+2J_1 + 3J_2} = 1 + (J_1^2 + 7J_1 J_2 +2J_2^2) - (2J_1^3 +13J_1^2 J_2 +17 J_1 J_2^2 + 8 J_2^3) \ .
\eea 
Note that $c_1(X)$ vanishes and $c_2(X) = c_2(\CA) = 50 \nu^1 + 44 \nu^2$, where $\nu^1,~ \nu^2$ are the 4-form basis elements satisfying 
\be \nn \int_X J_r \wedge \nu^s = \delta_r^s \ . \ee

\paragraph{K\"ahler cone:}
Our final task is to compute the K\"ahler and the Mori cone of $\CA$. 
Because the two cones are dual to each other, it is enough to work out the former. 
We first need to find all of the primitive collections, and, as can be seen in Figure \ref{f:palp}, PALP computes these as $\mathcal P_1 = 010001$ and $\mathcal P_2=101110$. Now, applying the inequality \eqref{strict} of Theorem \ref{kahler} to $\mathcal P_1$, the strictly-convexness condition becomes
\be \label{1}
\psi(\bold v_1 + \bold v_5) > \psi(\bold v_1) + \psi(\bold v_5) ~\Rightarrow~ 0 > -a^1 - a^5 \ , 
\ee
where $\psi(\bold v_\rho) = - a^\rho$ is obvious from the definition of support function.
Similarly, we have from the other primitive collection $\mathcal P_2$, 
\be \label{2}
\psi(\bold v_2 + \bold v_3 + \bold v_4 + \bold v_6) > \psi(\bold v_2) + \psi(\bold v_3) + \psi(\bold v_4) + \psi(\bold v_6) ~\Rightarrow~ -a^1 > -a^2 - a^3 - a^4 - a^6 \ .
\ee
As the K\"ahler cone lives in the vector space $H^{1,1}$, we had better express \eqref{1} and \eqref{2} in terms of $t^1 = \beta^1_{~\rho} a^\rho = a^1+a^5$ and $t^2=\beta^2_{~\rho} a^\rho = -a^1 + a^2 + a^3 + a^4 + a^6$. 
It is obvious to see that they are equivalent to 
\be
t^1 >0 ; ~ t^2>0 \ ,
\ee
which is exactly the first quadrant.

\begin{table}[thb]
{\begin{center}
\begin{tabular}{|@{}c@{}|c||@{}c@{}|c||@{}c@{}|c||@{}c@{}|c|}\hline 
no. & Vertices of $\Delta^\circ$  & no. &  Vertices of $\Delta^\circ$ & no. &  Vertices of $\Delta^\circ$ &  no. & Vertices of $\Delta^\circ$ \\ \hline \hline
\begin{tabular}{c}\small $1_P$\\ {\tiny $(1,101)$}\end{tabular}& 
\tiny $\begin{array}{cccc}
 1 & 0 & 0 & 0 \\
 0 & 1 & 0 & 0 \\
 0 & 0 & 1 & 0 \\
 0 & 0 & 0 & 1 \\
 -1 & -1 & -1 & -1
\end{array}$ 
&\begin{tabular}{c}\small $2_P$ \\ {\tiny $(2,86)$}\end{tabular}& 
\tiny $\begin{array}{cccc}
 1 & 0 & 0 & 0 \\
 0 & 1 & 0 & 0 \\
 0 & 0 & 1 & 0 \\
 0 & 0 & 0 & 1 \\
 -1 & 0 & 0 & 0 \\
 0 & -1 & -1 & -1
\end{array}$
&\begin{tabular}{c}\small 3\\ {\tiny $(2,90)$}\end{tabular}& 
\tiny $\begin{array}{cccc}
 1 & 0 & 0 & 0 \\
 0 & 1 & 0 & 0 \\
 0 & 0 & 1 & 0 \\
 0 & 0 & 0 & 1 \\
 -1 & 0 & 0 & 0 \\
 1 & -1 & -1 & -1
\end{array}$
& \begin{tabular}{c}\small 4 \\ {\tiny $(2,86)$}\end{tabular}& 
\tiny $\begin {array} {cccc}
 1 & 0 & 0 & 0 \\
 0 & 1 & 0 & 0 \\
 0 & 0 & 1 & 0 \\
 0 & 0 & 0 & 1 \\
 -1 & - 1 & 0 & 0 \\
 1 & 0 & - 1 & - 1
\end {array}$ \\ \hline
\begin{tabular}{c}\small5 \\ {\tiny $(2,86)$}\end{tabular}&
\tiny $\begin {array} {cccc}
 1 & 0 & 0 & 0 \\
 0 & 1 & 0 & 0 \\
 0 & 0 & 1 & 0 \\
 0 & 0 & 0 & 1 \\
 -1 & - 1 & 0 & 0 \\
 1 & 1 & - 1 & - 1
\end {array} $
& \begin{tabular}{c}\small6 \\ {\tiny $(2,86)$}\end{tabular}&
\tiny $\begin {array} {cccc}
 1 & 0 & 0 & 0 \\
 0 & 1 & 0 & 0 \\
 0 & 0 & 1 & 0 \\
 0 & 0 & 0 & 1 \\
 -1 & - 1 & - 1 & 0 \\
 1 & 0 & 0 & - 1
\end {array} $
& \begin{tabular}{c}\small $7_P$ \\ {\tiny $(2,83)$}\end{tabular}&
\tiny $\begin {array} {cccc}
 1 & 0 & 0 & 0 \\
 0 & 1 & 0 & 0 \\
 0 & 0 & 1 & 0 \\
 0 & 0 & 0 & 1 \\
 0 & - 1 & - 1 & 0 \\
 -1 & 0 & 0 & - 1
\end {array} $
& \begin{tabular}{c}\small8 \\ {\tiny $(2,102)$}\end{tabular}&
\tiny $\begin {array} {cccc}
 1 & 0 & 0 & 0 \\
 0 & 1 & 0 & 0 \\
 0 & 0 & 1 & 0 \\
 0 & 0 & 0 & 1 \\
 -1 & 0 & 0 & 0 \\
 2 & - 1 & - 1 & - 1
\end {array} $\\ \hline
\begin{tabular}{c}\small9 \\ {\tiny $(2,95)$}\end{tabular}&
\tiny $\begin {array} {cccc}
 1 & 0 & 0 & 0 \\
 0 & 1 & 0 & 0 \\
 0 & 0 & 1 & 0 \\
 0 & 0 & 0 & 1 \\
 -1 & - 1 & 0 & 0 \\
 2 & 0 & - 1 & - 1
\end {array} $
& \begin{tabular}{c}\small10 \\ {\tiny $(2,122)$}\end{tabular}&
\tiny $\begin {array} {cccc}
 1 & 0 & 0 & 0 \\
 0 & 1 & 0 & 0 \\
 0 & 0 & 1 & 0 \\
 0 & 0 & 0 & 1 \\
 -1 & 0 & 0 & 0 \\
 3 & - 1 & - 1 & - 1
\end {array} $
& \begin{tabular}{c}\small11 \\ {\tiny $(3,71)$}\end{tabular}&
\tiny $\begin {array} {cccc}
 1 & 0 & 0 & 0 \\
 0 & 1 & 0 & 0 \\
 0 & 0 & 1 & 0 \\
 0 & 0 & 0 & 1 \\
 -1 & 0 & 0 & 0 \\
 -1 & - 1 & 0 & 0 \\
 1 & 0 & - 1 & - 1
\end {array} $
& \begin{tabular}{c}\small12 \\ {\tiny $(3,75)$}\end{tabular}&
\tiny $\begin {array} {cccc}
 1 & 0 & 0 & 0 \\
 0 & 1 & 0 & 0 \\
 0 & 0 & 1 & 0 \\
 0 & 0 & 0 & 1 \\
 -1 & 0 & 0 & 0 \\
 0 & - 1 & - 1 & 0 \\
 0 & 1 & 0 & - 1
\end {array} $\\ \hline
\begin{tabular}{c}\small13 \\ {\tiny $(3,75)$}\end{tabular}&
\tiny $\begin {array} {cccc}
 1 & 0 & 0 & 0 \\
 0 & 1 & 0 & 0 \\
 0 & 0 & 1 & 0 \\
 0 & 0 & 0 & 1 \\
 -1 & 0 & 0 & 0 \\
 0 & - 1 & - 1 & 0 \\
 1 & 1 & 1 & - 1
\end {array} $
& \begin{tabular}{c}\small14 \\ {\tiny $(3,75)$}\end{tabular}&
\tiny $\begin {array} {cccc}
 1 & 0 & 0 & 0 \\
 0 & 1 & 0 & 0 \\
 0 & 0 & 1 & 0 \\
 0 & 0 & 0 & 1 \\
 -1 & 0 & 0 & 0 \\
 0 & - 1 & - 1 & - 1 \\
         1 & 1 & 0 & 0
\end {array} $
& \begin{tabular}{c}\small15 \\ {\tiny $(3,83)$}\end{tabular}&
\tiny $\begin {array} {cccc}
 1 & 0 & 0 & 0 \\
 0 & 1 & 0 & 0 \\
 0 & 0 & 1 & 0 \\
 0 & 0 & 0 & 1 \\
 -1 & 0 & 0 & 0 \\
 1 & - 1 & - 1 & - 1 \\
         1 & 1 & 0 & 0
\end {array} $
&\begin{tabular}{c}\small 16 \\ {\tiny $(3,79)$}\end{tabular}&
\tiny $\begin {array} {cccc}
 1 & 0 & 0 & 0 \\
 0 & 1 & 0 & 0 \\
 0 & 0 & 1 & 0 \\
 0 & 0 & 0 & 1 \\
 -1 & 0 & 0 & 0 \\
 1 & - 1 & - 1 & 0 \\
 0 & 1 & 0 & - 1
\end {array} $\\ \hline
\begin{tabular}{c}\small17 \\ {\tiny $(3,75)$}\end{tabular}&
\tiny $\begin {array} {cccc}
 1 & 0 & 0 & 0 \\
 0 & 1 & 0 & 0 \\
 0 & 0 & 1 & 0 \\
 0 & 0 & 0 & 1 \\
 -1 & 0 & 0 & 0 \\
 1 & 0 & - 1 & 0 \\
 0 & - 1 & 0 & - 1
\end {array} $
&\begin{tabular}{c}\small 18 \\ {\tiny $(3,87)$}\end{tabular}&
\tiny $\begin {array} {cccc}
 1 & 0 & 0 & 0 \\
 0 & 1 & 0 & 0 \\
 0 & 0 & 1 & 0 \\
 0 & 0 & 0 & 1 \\
 -1 & 0 & 0 & 0 \\
 1 & - 1 & - 1 & 0 \\
 1 & 1 & 1 & - 1
\end {array} $
& \begin{tabular}{c}\small19 \\ {\tiny $(3,75)$}\end{tabular}&
\tiny $\begin {array} {cccc}
 1 & 0 & 0 & 0 \\
 0 & 1 & 0 & 0 \\
 0 & 0 & 1 & 0 \\
 0 & 0 & 0 & 1 \\
 -1 & 0 & 0 & 0 \\
 1 & - 1 & 0 & 0 \\
 -1 & 1 & - 1 & - 1
\end {array} $
&\begin{tabular}{c}\small 20 \\ {\tiny $(3,81)$}\end{tabular}&
\tiny $\begin {array} {cccc}
 1 & 0 & 0 & 0 \\
 0 & 1 & 0 & 0 \\
 0 & 0 & 1 & 0 \\
 0 & 0 & 0 & 1 \\
 -1 & 0 & 0 & 0 \\
 1 & - 1 & 0 & 0 \\
 0 & 1 & - 1 & - 1
\end {array} $\\ \hline
\begin{tabular}{c}\small21 \\ {\tiny $(3,83)$}\end{tabular}&
\tiny $\begin {array} {cccc}
 1 & 0 & 0 & 0 \\
 0 & 1 & 0 & 0 \\
 0 & 0 & 1 & 0 \\
 0 & 0 & 0 & 1 \\
 -1 & 0 & 0 & 0 \\
 1 & - 1 & 0 & 0 \\
 1 & 0 & - 1 & - 1
\end {array} $
& \begin{tabular}{c}\small22 \\ {\tiny $(3,72)$}\end{tabular}&
\tiny $\begin {array} {cccc}
 1 & 0 & 0 & 0 \\
 0 & 1 & 0 & 0 \\
 0 & 0 & 1 & 0 \\
 0 & 0 & 0 & 1 \\
 -1 & - 1 & 0 & 0 \\
 0 & 1 & - 1 & 0 \\
 1 & 0 & 0 & - 1
\end {array} $
& \begin{tabular}{c}\small23 \\ {\tiny $(3,73)$}\end{tabular}&
\tiny $\begin {array} {cccc}
 1 & 0 & 0 & 0 \\
 0 & 1 & 0 & 0 \\
 0 & 0 & 1 & 0 \\
 0 & 0 & 0 & 1 \\
 -1 & - 1 & 0 & 0 \\
 1 & 1 & - 1 & - 1 \\
         0 & - 1 & 1 & 0
\end {array} $
& \begin{tabular}{c}\small24 \\ {\tiny $(3,78)$}\end{tabular}&
\tiny $\begin {array} {cccc}
 1 & 0 & 0 & 0 \\
 0 & 1 & 0 & 0 \\
 0 & 0 & 1 & 0 \\
 0 & 0 & 0 & 1 \\
 -1 & - 1 & 0 & 0 \\
 1 & 0 & - 1 & 0 \\
 0 & 0 & 1 & - 1
\end {array} $\\ \hline
\begin{tabular}{c}\small25 \\ {\tiny $(3,81)$}\end{tabular}&
\tiny $\begin {array} {cccc}
 1 & 0 & 0 & 0 \\
 0 & 1 & 0 & 0 \\
 0 & 0 & 1 & 0 \\
 0 & 0 & 0 & 1 \\
 -1 & - 1 & 0 & 0 \\
 1 & 0 & - 1 & 0 \\
 1 & 0 & 0 & - 1
\end {array} $
& \begin{tabular}{c}\small $26_P$ \\ {\tiny $(3,75)$}\end{tabular}&
\tiny $\begin {array} {cccc}
 1 & 0 & 0 & 0 \\
 0 & 1 & 0 & 0 \\
 0 & 0 & 1 & 0 \\
 0 & 0 & 0 & 1 \\
 0 & - 1 & 0 & 0 \\
 -1 & 0 & 0 & 0 \\
 0 & 0 & - 1 & - 1
\end {array} $
& \begin{tabular}{c}\small27 \\ {\tiny $(3,77)$}\end{tabular}&
\tiny $\begin {array} {cccc}
 1 & 0 & 0 & 0 \\
 0 & 1 & 0 & 0 \\
 0 & 0 & 1 & 0 \\
 0 & 0 & 0 & 1 \\
 0 & - 1 & 0 & 0 \\
 -1 & 0 & 0 & 0 \\
 1 & 0 & - 1 & - 1
\end {array} $
& \begin{tabular}{c}\small28 \\ {\tiny $(3,79)$}\end{tabular}&
\tiny $\begin {array} {cccc}
 1 & 0 & 0 & 0 \\
 0 & 1 & 0 & 0 \\
 0 & 0 & 1 & 0 \\
 0 & 0 & 0 & 1 \\
 0 & - 1 & 0 & 0 \\
 -1 & 0 & 0 & 0 \\
 1 & 1 & - 1 & - 1
\end {array} $\\ \hline
\begin{tabular}{c}\small29 \\ {\tiny $(3,71)$}\end{tabular}&
\tiny $\begin {array} {cccc}
 1 & 0 & 0 & 0 \\
 0 & 1 & 0 & 0 \\
 0 & 0 & 1 & 0 \\
 0 & 0 & 0 & 1 \\
 0 & - 1 & - 1 & 0 \\
 -1 & 0 & 0 & - 1 \\
         1 & 1 & 0 & 0
\end {array} $
& \begin{tabular}{c}\small30 \\ {\tiny $(3,72)$}\end{tabular}&
\tiny $\begin {array} {cccc}
 1 & 0 & 0 & 0 \\
 0 & 1 & 0 & 0 \\
 0 & 0 & 1 & 0 \\
 0 & 0 & 0 & 1 \\
 0 & - 1 & - 1 & 0 \\
 -1 & 0 & 0 & - 1 \\
         1 & 1 & 1 & 0
\end {array} $
&\begin{tabular}{c}\small 31 \\ {\tiny $(3,71)$}\end{tabular}&
\tiny $\begin {array} {cccc}
 1 & 0 & 0 & 0 \\
 0 & 1 & 0 & 0 \\
 0 & 0 & 1 & 0 \\
 0 & 0 & 0 & 1 \\
 -1 & 0 & 0 & 0 \\
 -1 & - 1 & 0 & 0 \\
 2 & 0 & - 1 & - 1
\end {array} $
& \begin{tabular}{c}\small32 \\ {\tiny $(3,99)$}\end{tabular}&
\tiny $\begin {array} {cccc}
 1 & 0 & 0 & 0 \\
 0 & 1 & 0 & 0 \\
 0 & 0 & 1 & 0 \\
 0 & 0 & 0 & 1 \\
 -1 & 0 & 0 & 0 \\
 2 & - 1 & - 1 & - 1 \\
         1 & 1 & 0 & 0
\end {array} $\\ \hline
\begin{tabular}{c}\small33 \\ {\tiny $(3,91)$}\end{tabular}&
\tiny $\begin {array} {cccc}
 1 & 0 & 0 & 0 \\
 0 & 1 & 0 & 0 \\
 0 & 0 & 1 & 0 \\
 0 & 0 & 0 & 1 \\
 -1 & 0 & 0 & 0 \\
 2 & - 1 & - 1 & 0 \\
 0 & 1 & 0 & - 1
\end {array} $
&\begin{tabular}{c}\small 34 \\ {\tiny $(3,93)$}\end{tabular}&
\tiny $\begin {array} {cccc}
 1 & 0 & 0 & 0 \\
 0 & 1 & 0 & 0 \\
 0 & 0 & 1 & 0 \\
 0 & 0 & 0 & 1 \\
 -1 & 0 & 0 & 0 \\
 1 & - 1 & 0 & 0 \\
 0 & 2 & - 1 & - 1
\end {array} $
&\begin{tabular}{c}\small 35 \\ {\tiny $(3,91)$}\end{tabular}&
\tiny $\begin {array} {cccc}
 1 & 0 & 0 & 0 \\
 0 & 1 & 0 & 0 \\
 0 & 0 & 1 & 0 \\
 0 & 0 & 0 & 1 \\
 -1 & 0 & 0 & 0 \\
 1 & - 1 & 0 & 0 \\
 2 & - 1 & - 1 & - 1
\end {array} $
&\begin{tabular}{c}\small 36 \\ {\tiny $(3,95)$}\end{tabular}&
\tiny $\begin {array} {cccc}
 1 & 0 & 0 & 0 \\
 0 & 1 & 0 & 0 \\
 0 & 0 & 1 & 0 \\
 0 & 0 & 0 & 1 \\
 -1 & 0 & 0 & 0 \\
 1 & - 1 & 0 & 0 \\
 2 & 0 & - 1 & - 1
\end {array} $\\ \hline
\begin{tabular}{c}\small37 \\ {\tiny $(3,76)$}\end{tabular}&
\tiny $\begin {array} {cccc}
 1 & 0 & 0 & 0 \\
 0 & 1 & 0 & 0 \\
 0 & 0 & 1 & 0 \\
 0 & 0 & 0 & 1 \\
 -1 & - 1 & 0 & 0 \\
 1 & 0 & - 1 & - 1 \\
         2 & 0 & - 1 & - 1
\end {array} $
& \begin{tabular}{c}\small38 \\ {\tiny $(3,83)$}\end{tabular}&
\tiny $\begin {array} {cccc}
 1 & 0 & 0 & 0 \\
 0 & 1 & 0 & 0 \\
 0 & 0 & 1 & 0 \\
 0 & 0 & 0 & 1 \\
 0 & - 1 & 0 & 0 \\
 -1 & 0 & 0 & 0 \\
 2 & 0 & - 1 & - 1
\end {array} $
&\begin{tabular}{c}\small 39 \\ {\tiny $(4,60)$}\end{tabular}&
\tiny $\begin {array} {cccc}
 1 & 0 & 0 & 0 \\
 0 & 1 & 0 & 0 \\
 0 & 0 & 1 & 0 \\
 0 & 0 & 0 & 1 \\
 0 & - 1 & - 1 & 0 \\
 -1 & 0 & 0 & - 1 \\
         1 & 1 & 0 & 0 \\
 0 & 0 & 1 & 1
\end {array} $
&\begin{tabular}{c}\small $40_P$ \\ {\tiny $(4,68)$}\end{tabular}&
\tiny $\begin {array} {cccc}
 1 & 0 & 0 & 0 \\
 0 & 1 & 0 & 0 \\
 0 & 0 & 1 & 0 \\
 0 & 0 & 0 & 1 \\
 0 & 0 & 0 & - 1 \\
 0 & 0 & - 1 & 0 \\
 0 & - 1 & 0 & 0 \\
 -1 & 0 & 0 & 0
\end {array} $\\ \hline
\end{tabular}
\end{center}}
\end{table}
\begin{table}[thb]
{\begin{center}
\begin{tabular}{|@{}c@{}|c||@{}c@{}|c||@{}c@{}|c||@{}c@{}|c|}\hline 
no. & Vertices of $\Delta^\circ$  & no. &  Vertices of $\Delta^\circ$ & no. &  Vertices of $\Delta^\circ$ &  no. & Vertices of $\Delta^\circ$ \\ \hline \hline
\begin{tabular}{c}\small41 \\ {\tiny $(4,63)$}\end{tabular}&
\tiny $\begin {array} {cccc}
 1 & 0 & 0 & 0 \\
 0 & 1 & 0 & 0 \\
 0 & 0 & 1 & 0 \\
 0 & 0 & 0 & 1 \\
 -1 & 0 & 0 & 0 \\
 -1 & - 1 & 0 & 0 \\
 0 & 0 & - 1 & - 1 \\
         1 & 0 & 1 & 0
\end {array} $
&\begin{tabular}{c}\small 42 \\ {\tiny $(4,61)$}\end{tabular}&
\tiny $\begin {array} {cccc}
 1 & 0 & 0 & 0 \\
 0 & 1 & 0 & 0 \\
 0 & 0 & 1 & 0 \\
 0 & 0 & 0 & 1 \\
 -1 & 0 & 0 & 0 \\
 -1 & - 1 & 0 & 0 \\
 1 & 0 & - 1 & - 1 \\
         1 & 0 & 1 & 0
\end {array} $
& \begin{tabular}{c}\small43 \\ {\tiny $(4,64)$}\end{tabular}&
\tiny $\begin {array} {cccc}
 1 & 0 & 0 & 0 \\
 0 & 1 & 0 & 0 \\
 0 & 0 & 1 & 0 \\
 0 & 0 & 0 & 1 \\
 -1 & 0 & 0 & 0 \\
 -1 & - 1 & 0 & 0 \\
 1 & 0 & - 1 & 0 \\
 0 & 0 & 1 & - 1
\end {array} $
& \begin{tabular}{c}\small $44_N$ \\ {\tiny $(4,61)$}\end{tabular}&
\tiny $\begin {array} {cccc}
 1 & 0 & 0 & 0 \\
 0 & 1 & 0 & 0 \\
 0 & 0 & 1 & 0 \\
 0 & 0 & 0 & 1 \\
 -1 & 0 & 0 & 0 \\
 -1 & 1 & 0 & 0 \\
 1 & - 1 & - 1 & 0 \\
 0 & - 1 & 0 & - 1
\end {array} $\\ \hline
\begin{tabular}{c}\small45 \\ {\tiny $(4,64)$}\end{tabular}&
\tiny $\begin {array} {cccc}
 1 & 0 & 0 & 0 \\
 0 & 1 & 0 & 0 \\
 0 & 0 & 1 & 0 \\
 0 & 0 & 0 & 1 \\
 -1 & 0 & 0 & 0 \\
 -1 & 0 & - 1 & 0 \\
 1 & 0 & 0 & - 1 \\
         1 & - 1 & 0 & 0
\end {array} $
& \begin{tabular}{c}\small46 \\ {\tiny $(4,68)$}\end{tabular}&
\tiny $\begin {array} {cccc}
 1 & 0 & 0 & 0 \\
 0 & 1 & 0 & 0 \\
 0 & 0 & 1 & 0 \\
 0 & 0 & 0 & 1 \\
 -1 & 0 & 0 & 0 \\
 0 & - 1 & - 1 & 0 \\
 1 & 1 & 0 & 0 \\
 0 & 1 & 0 & - 1
\end {array} $
& \begin{tabular}{c}\small47 \\ {\tiny $(4,66)$}\end{tabular}&
\tiny $\begin {array} {cccc}
 1 & 0 & 0 & 0 \\
 0 & 1 & 0 & 0 \\
 0 & 0 & 1 & 0 \\
 0 & 0 & 0 & 1 \\
 -1 & 0 & 0 & 0 \\
 0 & - 1 & - 1 & 0 \\
 1 & 0 & 0 & - 1 \\
         0 & 1 & 1 & 1
\end {array} $
& \begin{tabular}{c}\small48 \\ {\tiny $(4,65)$}\end{tabular}&
\tiny $\begin {array} {cccc}
 1 & 0 & 0 & 0 \\
 0 & 1 & 0 & 0 \\
 0 & 0 & 1 & 0 \\
 0 & 0 & 0 & 1 \\
 -1 & 0 & 0 & 0 \\
 0 & - 1 & - 1 & 0 \\
 1 & 1 & 0 & 0 \\
 0 & 0 & 1 & - 1
\end {array} $\\ \hline
\begin{tabular}{c}\small $49_N$ \\ {\tiny $(4,67)$}\end{tabular}&
\tiny $\begin {array} {cccc}
 1 & 0 & 0 & 0 \\
 0 & 1 & 0 & 0 \\
 0 & 0 & 1 & 0 \\
 0 & 0 & 0 & 1 \\
 -1 & 0 & 0 & 0 \\
 0 & - 1 & - 1 & 0 \\
 1 & 1 & 0 & 0 \\
 1 & - 1 & 0 & - 1
\end {array} $
& \begin{tabular}{c}\small50 \\ {\tiny $(4,71)$}\end{tabular}&
\tiny $\begin {array} {cccc}
 1 & 0 & 0 & 0 \\
 0 & 1 & 0 & 0 \\
 0 & 0 & 1 & 0 \\
 0 & 0 & 0 & 1 \\
 -1 & 0 & 0 & 0 \\
 0 & - 1 & - 1 & 0 \\
 1 & 1 & 0 & 0 \\
 1 & 1 & 0 & - 1
\end {array} $
& \begin{tabular}{c}\small51 \\ {\tiny $(4,73)$}\end{tabular}&
\tiny $\begin {array} {cccc}
 1 & 0 & 0 & 0 \\
 0 & 1 & 0 & 0 \\
 0 & 0 & 1 & 0 \\
 0 & 0 & 0 & 1 \\
 -1 & 0 & 0 & 0 \\
 1 & - 1 & - 1 & 0 \\
 1 & 1 & 0 & 0 \\
 0 & 1 & 0 & - 1
\end {array} $
& \begin{tabular}{c}\small52 \\ {\tiny $(4,80)$}\end{tabular}&
\tiny $\begin {array} {cccc}
 1 & 0 & 0 & 0 \\
 0 & 1 & 0 & 0 \\
 0 & 0 & 1 & 0 \\
 0 & 0 & 0 & 1 \\
 -1 & 0 & 0 & 0 \\
 1 & - 1 & 0 & 0 \\
 1 & 0 & 0 & - 1 \\
         1 & 0 & - 1 & 0
\end {array} $\\ \hline
\begin{tabular}{c}\small53 \\ {\tiny $(4,72)$}\end{tabular}&
\tiny $\begin {array} {cccc}
 1 & 0 & 0 & 0 \\
 0 & 1 & 0 & 0 \\
 0 & 0 & 1 & 0 \\
 0 & 0 & 0 & 1 \\
 -1 & 0 & 0 & 0 \\
 1 & - 1 & 0 & 0 \\
 1 & 0 & - 1 & 0 \\
 -1 & 1 & 1 & - 1
\end {array} $
& \begin{tabular}{c}\small54 \\ {\tiny $(4,73)$}\end{tabular}&
\tiny $\begin {array} {cccc}
 1 & 0 & 0 & 0 \\
 0 & 1 & 0 & 0 \\
 0 & 0 & 1 & 0 \\
 0 & 0 & 0 & 1 \\
 -1 & 0 & 0 & 0 \\
 1 & - 1 & - 1 & 0 \\
 1 & 1 & 0 & 0 \\
 0 & 0 & 1 & - 1
\end {array} $
& \begin{tabular}{c}\small55 \\ {\tiny $(4,82)$}\end{tabular}&
\tiny $\begin {array} {cccc}
 1 & 0 & 0 & 0 \\
 0 & 1 & 0 & 0 \\
 0 & 0 & 1 & 0 \\
 0 & 0 & 0 & 1 \\
 -1 & 0 & 0 & 0 \\
 1 & - 1 & - 1 & 0 \\
 1 & 1 & 0 & 0 \\
 1 & 1 & 0 & - 1
\end {array} $
&\begin{tabular}{c}\small 56 \\ {\tiny $(4,68)$}\end{tabular}&
\tiny $\begin {array} {cccc}
 1 & 0 & 0 & 0 \\
 0 & 1 & 0 & 0 \\
 0 & 0 & 1 & 0 \\
 0 & 0 & 0 & 1 \\
 -1 & 0 & 0 & 0 \\
 1 & - 1 & 0 & 0 \\
 -1 & 1 & - 1 & - 1 \\
         1 & 0 & 1 & 0
\end {array} $\\ \hline
\begin{tabular}{c}\small57 \\ {\tiny $(4,74)$}\end{tabular}&
\tiny $\begin {array} {cccc}
 1 & 0 & 0 & 0 \\
 0 & 1 & 0 & 0 \\
 0 & 0 & 1 & 0 \\
 0 & 0 & 0 & 1 \\
 -1 & 0 & 0 & 0 \\
 1 & - 1 & 0 & 0 \\
 0 & 1 & - 1 & 0 \\
 0 & 0 & 1 & - 1
\end {array} $
& \begin{tabular}{c}\small58 \\ {\tiny $(4,78)$}\end{tabular}&
\tiny $\begin {array} {cccc}
 1 & 0 & 0 & 0 \\
 0 & 1 & 0 & 0 \\
 0 & 0 & 1 & 0 \\
 0 & 0 & 0 & 1 \\
 -1 & 0 & 0 & 0 \\
 1 & - 1 & 0 & 0 \\
 0 & 1 & - 1 & 0 \\
 0 & 1 & 0 & - 1
\end {array} $
& \begin{tabular}{c}\small59 \\ {\tiny $(4,69)$}\end{tabular}&
\tiny $\begin {array} {cccc}
 1 & 0 & 0 & 0 \\
 0 & 1 & 0 & 0 \\
 0 & 0 & 1 & 0 \\
 0 & 0 & 0 & 1 \\
 -1 & 0 & 0 & 0 \\
 1 & - 1 & 0 & 0 \\
 0 & 0 & - 1 & - 1 \\
         1 & 0 & 1 & 0
\end {array} $
& \begin{tabular}{c}\small60 \\ {\tiny $(4,76)$}\end{tabular}&
\tiny $\begin {array} {cccc}
 1 & 0 & 0 & 0 \\
 0 & 1 & 0 & 0 \\
 0 & 0 & 1 & 0 \\
 0 & 0 & 0 & 1 \\
 -1 & 0 & 0 & 0 \\
 1 & - 1 & 0 & 0 \\
 0 & 1 & - 1 & - 1 \\
         1 & 0 & 1 & 0
\end {array} $\\ \hline
\begin{tabular}{c}\small61 \\ {\tiny $(4,68)$}\end{tabular}&
\tiny $\begin {array} {cccc}
 1 & 0 & 0 & 0 \\
 0 & 1 & 0 & 0 \\
 0 & 0 & 1 & 0 \\
 0 & 0 & 0 & 1 \\
 -1 & 0 & 0 & 0 \\
 1 & - 1 & 0 & 0 \\
 1 & - 1 & - 1 & 0 \\
 0 & 1 & 0 & - 1
\end {array} $
& \begin{tabular}{c}\small62 \\ {\tiny $(4,79)$}\end{tabular}&
\tiny $\begin {array} {cccc}
 1 & 0 & 0 & 0 \\
 0 & 1 & 0 & 0 \\
 0 & 0 & 1 & 0 \\
 0 & 0 & 0 & 1 \\
 -1 & 0 & 0 & 0 \\
 1 & - 1 & 0 & 0 \\
 1 & 0 & - 1 & - 1 \\
         1 & 0 & 1 & 0
\end {array} $
&\begin{tabular}{c}\small 63 \\ {\tiny $(4,76)$}\end{tabular}&
\tiny $\begin {array} {cccc}
 1 & 0 & 0 & 0 \\
 0 & 1 & 0 & 0 \\
 0 & 0 & 1 & 0 \\
 0 & 0 & 0 & 1 \\
 -1 & 0 & 0 & 0 \\
 1 & - 1 & 0 & 0 \\
 1 & 0 & - 1 & 0 \\
 0 & 0 & 1 & - 1
\end {array} $
& \begin{tabular}{c}\small64 \\ {\tiny $(4,64)$}\end{tabular}&
\tiny $\begin {array} {cccc}
 1 & 0 & 0 & 0 \\
 0 & 1 & 0 & 0 \\
 0 & 0 & 1 & 0 \\
 0 & 0 & 0 & 1 \\
 0 & - 1 & 0 & 0 \\
 -1 & 0 & 0 & 0 \\
 -1 & - 1 & 0 & 0 \\
 1 & 1 & - 1 & - 1
\end {array} $\\ \hline
\begin{tabular}{c}\small65 \\ {\tiny $(4,66)$}\end{tabular}&
\tiny $\begin {array} {cccc}
 1 & 0 & 0 & 0 \\
 0 & 1 & 0 & 0 \\
 0 & 0 & 1 & 0 \\
 0 & 0 & 0 & 1 \\
 0 & - 1 & 0 & 0 \\
 -1 & 0 & 0 & 0 \\
 -1 & 1 & 0 & 0 \\
 1 & 0 & - 1 & - 1
\end {array} $
&\begin{tabular}{c}\small 66 \\ {\tiny $(4,72)$}\end{tabular}&
\tiny $\begin {array} {cccc}
 1 & 0 & 0 & 0 \\
 0 & 1 & 0 & 0 \\
 0 & 0 & 1 & 0 \\
 0 & 0 & 0 & 1 \\
 0 & - 1 & 0 & 0 \\
 -1 & 0 & 0 & 0 \\
 -1 & 1 & 0 & 0 \\
 1 & 1 & - 1 & - 1
\end {array} $
&\begin{tabular}{c}\small 67 \\ {\tiny $(4,64)$}\end{tabular}&
\tiny $\begin {array} {cccc}
 1 & 0 & 0 & 0 \\
 0 & 1 & 0 & 0 \\
 0 & 0 & 1 & 0 \\
 0 & 0 & 0 & 1 \\
 0 & - 1 & 0 & 0 \\
 -1 & 0 & 0 & 0 \\
 -1 & 0 & - 1 & 0 \\
 1 & 0 & 0 & - 1
\end {array} $
&\begin{tabular}{c}\small 68 \\ {\tiny $(4,66)$}\end{tabular}&
\tiny $\begin {array} {cccc}
 1 & 0 & 0 & 0 \\
 0 & 1 & 0 & 0 \\
 0 & 0 & 1 & 0 \\
 0 & 0 & 0 & 1 \\
 0 & - 1 & 0 & 0 \\
 -1 & 0 & 0 & 0 \\
 0 & 0 & - 1 & - 1 \\
         1 & 0 & 1 & 0
\end {array} $\\ \hline
\begin{tabular}{c}\small69 \\ {\tiny $(4,68)$}\end{tabular}&
\tiny $\begin {array} {cccc}
 1 & 0 & 0 & 0 \\
 0 & 1 & 0 & 0 \\
 0 & 0 & 1 & 0 \\
 0 & 0 & 0 & 1 \\
 0 & - 1 & 0 & 0 \\
 -1 & 0 & 0 & 0 \\
 0 & 1 & 0 & - 1 \\
         1 & 0 & - 1 & 0
\end {array} $
&\begin{tabular}{c}\small 70 \\ {\tiny $(4,66)$}\end{tabular}&
\tiny $\begin {array} {cccc}
 1 & 0 & 0 & 0 \\
 0 & 1 & 0 & 0 \\
 0 & 0 & 1 & 0 \\
 0 & 0 & 0 & 1 \\
 0 & - 1 & 0 & 0 \\
 -1 & 0 & 0 & 0 \\
 0 & 1 & - 1 & 0 \\
 1 & 0 & - 1 & - 1
\end {array} $
&\begin{tabular}{c}\small 71 \\ {\tiny $(4,68)$}\end{tabular}&
\tiny $\begin {array} {cccc}
 1 & 0 & 0 & 0 \\
 0 & 1 & 0 & 0 \\
 0 & 0 & 1 & 0 \\
 0 & 0 & 0 & 1 \\
 0 & - 1 & 0 & 0 \\
 -1 & 0 & 0 & 0 \\
 1 & 0 & - 1 & - 1 \\
         0 & 1 & 1 & 0
\end {array} $
&\begin{tabular}{c}\small 72 \\ {\tiny $(4,70)$}\end{tabular}&
\tiny $\begin {array} {cccc}
 1 & 0 & 0 & 0 \\
 0 & 1 & 0 & 0 \\
 0 & 0 & 1 & 0 \\
 0 & 0 & 0 & 1 \\
 0 & - 1 & 0 & 0 \\
 -1 & 0 & 0 & 0 \\
 1 & 0 & - 1 & - 1 \\
         1 & 0 & 1 & 0
\end {array} $\\ \hline
\end{tabular}
\end{center}}
\end{table}
\begin{table}[thb]
{\begin{center}
\begin{tabular}{|@{}c@{}|c||@{}c@{}|c||@{}c@{}|c||@{}c@{}|c|}\hline no. & Vertices of $\Delta^\circ$  & no. &  Vertices of $\Delta^\circ$ & no. &  Vertices of $\Delta^\circ$ &  no. & Vertices of $\Delta^\circ$ \\ \hline \hline
\begin{tabular}{c}\small73 \\ {\tiny $(4,72)$}\end{tabular}&
\tiny $\begin {array} {cccc}
 1 & 0 & 0 & 0 \\
 0 & 1 & 0 & 0 \\
 0 & 0 & 1 & 0 \\
 0 & 0 & 0 & 1 \\
 0 & - 1 & 0 & 0 \\
 -1 & 0 & 0 & 0 \\
 1 & 0 & 0 & - 1 \\
         1 & 0 & - 1 & 0
\end {array} $
&\begin{tabular}{c}\small 74 \\ {\tiny $(4,70)$}\end{tabular}&
\tiny $\begin {array} {cccc}
 1 & 0 & 0 & 0 \\
 0 & 1 & 0 & 0 \\
 0 & 0 & 1 & 0 \\
 0 & 0 & 0 & 1 \\
 0 & - 1 & 0 & 0 \\
 -1 & 0 & 0 & 0 \\
 1 & 0 & - 1 & 0 \\
 0 & 0 & 1 & - 1
\end {array} $
&\begin{tabular}{c}\small 75 \\ {\tiny $(4,67)$}\end{tabular}&
\tiny $\begin {array} {cccc}
 1 & 0 & 0 & 0 \\
 0 & 1 & 0 & 0 \\
 0 & 0 & 1 & 0 \\
 0 & 0 & 0 & 1 \\
 0 & - 1 & 0 & 0 \\
 -1 & 0 & 0 & 0 \\
 1 & 1 & 0 & 0 \\
 0 & 0 & - 1 & - 1
\end {array} $
&\begin{tabular}{c}\small 76 \\ {\tiny $(4,72)$}\end{tabular}&
\tiny $\begin {array} {cccc}
 1 & 0 & 0 & 0 \\
 0 & 1 & 0 & 0 \\
 0 & 0 & 1 & 0 \\
 0 & 0 & 0 & 1 \\
 0 & - 1 & 0 & 0 \\
 -1 & 0 & 0 & 0 \\
 1 & 1 & 0 & 0 \\
 1 & 0 & - 1 & - 1
\end {array} $\\ \hline
\begin{tabular}{c}\small77 \\ {\tiny $(4,76)$}\end{tabular}&
\tiny $\begin {array} {cccc}
 1 & 0 & 0 & 0 \\
 0 & 1 & 0 & 0 \\
 0 & 0 & 1 & 0 \\
 0 & 0 & 0 & 1 \\
 0 & - 1 & 0 & 0 \\
 -1 & 0 & 0 & 0 \\
 1 & 1 & 0 & 0 \\
 1 & 1 & - 1 & - 1
\end {array} $
&\begin{tabular}{c}\small 78 \\ {\tiny $(4,68)$}\end{tabular}&
\tiny $\begin {array} {cccc}
 1 & 0 & 0 & 0 \\
 0 & 1 & 0 & 0 \\
 0 & 0 & 1 & 0 \\
 0 & 0 & 0 & 1 \\
 0 & 0 & - 1 & 0 \\
 0 & - 1 & 0 & 0 \\
 -1 & 0 & 0 & 0 \\
 1 & 0 & 0 & - 1
\end {array} $
&\begin{tabular}{c}\small 79 \\ {\tiny $(4,68)$}\end{tabular}&
\tiny $\begin {array} {cccc}
 1 & 0 & 0 & 0 \\
 0 & 1 & 0 & 0 \\
 0 & 0 & 1 & 0 \\
 0 & 0 & 0 & 1 \\
 0 & 0 & - 1 & 0 \\
 0 & - 1 & 0 & 0 \\
 -1 & 0 & 0 & 0 \\
 1 & 1 & 1 & - 1
\end {array} $
&\begin{tabular}{c}\small $80_N$ \\ {\tiny $(4,61)$}\end{tabular}&
\tiny $\begin {array} {cccc}
 1 & 0 & 0 & 0 \\
 0 & 1 & 0 & 0 \\
 0 & 0 & 1 & 0 \\
 0 & 0 & 0 & 1 \\
 0 & - 1 & - 1 & 0 \\
 -1 & 0 & 0 & - 1 \\
         1 & 1 & 1 & 0 \\
 0 & 0 & - 1 & 1
\end {array} $\\ \hline
\begin{tabular}{c}\small81 \\ {\tiny $(4,65)$}\end{tabular}&
\tiny $\begin {array} {cccc}
 1 & 0 & 0 & 0 \\
 0 & 1 & 0 & 0 \\
 0 & 0 & 1 & 0 \\
 0 & 0 & 0 & 1 \\
 0 & - 1 & - 1 & 0 \\
 -1 & 0 & 0 & - 1 \\
         1 & 1 & 1 & 0 \\
 1 & 1 & 0 & 0
\end {array} $
&\begin{tabular}{c}\small 82 \\ {\tiny $(4,69)$}\end{tabular}&
\tiny $\begin {array} {cccc}
 1 & 0 & 0 & 0 \\
 0 & 1 & 0 & 0 \\
 0 & 0 & 1 & 0 \\
 0 & 0 & 0 & 1 \\
 0 & - 1 & 0 & 0 \\
 -1 & 0 & 0 & 0 \\
 -1 & 1 & 0 & 0 \\
 2 & 0 & - 1 & - 1
\end {array} $
&\begin{tabular}{c}\small 83 \\ {\tiny $(4,81)$}\end{tabular}&
\tiny $\begin {array} {cccc}
 1 & 0 & 0 & 0 \\
 0 & 1 & 0 & 0 \\
 0 & 0 & 1 & 0 \\
 0 & 0 & 0 & 1 \\
 0 & - 1 & 0 & 0 \\
 -1 & 0 & 0 & 0 \\
 1 & 1 & 0 & 0 \\
 2 & 0 & - 1 & - 1
\end {array} $
&\begin{tabular}{c}\small 84 \\ {\tiny $(4,84)$}\end{tabular}&
\tiny $\begin {array} {cccc}
 1 & 0 & 0 & 0 \\
 0 & 1 & 0 & 0 \\
 0 & 0 & 1 & 0 \\
 0 & 0 & 0 & 1 \\
 0 & - 1 & 0 & 0 \\
 -1 & 0 & 0 & 0 \\
 1 & 1 & 0 & 0 \\
 2 & 1 & - 1 & - 1
\end {array} $\\ \hline
\begin{tabular}{c}\small85 \\ {\tiny $(4,91)$}\end{tabular}&
\tiny $\begin {array} {cccc}
 1 & 0 & 0 & 0 \\
 0 & 1 & 0 & 0 \\
 0 & 0 & 1 & 0 \\
 0 & 0 & 0 & 1 \\
 0 & - 1 & 0 & 0 \\
 -1 & 0 & 0 & 0 \\
 1 & 1 & 0 & 0 \\
 2 & 2 & - 1 & - 1
\end {array} $
& \begin{tabular}{c}\small86 \\ {\tiny $(5,61)$}\end{tabular}&
\tiny $\begin {array} {cccc}
 1 & 0 & 0 & 0 \\
 0 & 1 & 0 & 0 \\
 0 & 0 & 1 & 0 \\
 0 & 0 & 0 & 1 \\
 0 & 0 & 0 & - 1 \\
         0 & 0 & - 1 & 0 \\
 0 & - 1 & 0 & 0 \\
 -1 & 0 & 0 & 0 \\
 1 & 1 & 0 & 0
\end {array} $
&\begin{tabular}{c}\small $87_N$ \\ {\tiny $(5,57)$}\end{tabular}&
\tiny $\begin {array} {cccc}
 1 & 0 & 0 & 0 \\
 0 & 1 & 0 & 0 \\
 0 & 0 & 1 & 0 \\
 0 & 0 & 0 & 1 \\
 0 & 0 & 0 & - 1 \\
         0 & 0 & - 1 & 0 \\
 0 & - 1 & 0 & 0 \\
 -1 & 0 & 0 & 0 \\
 1 & 1 & 1 & 1
\end {array} $
& \begin{tabular}{c}\small $88_N$ \\ {\tiny $(5,59)$}\end{tabular}&
\tiny $\begin {array} {cccc}
 1 & 0 & 0 & 0 \\
 0 & 1 & 0 & 0 \\
 0 & 0 & 1 & 0 \\
 0 & 0 & 0 & 1 \\
 0 & - 1 & 0 & 0 \\
 -1 & 0 & 0 & 0 \\
 -1 & 1 & 0 & 0 \\
 0 & 1 & - 1 & 0 \\
 1 & - 1 & 1 & - 1
\end {array} $\\ \hline
\begin{tabular}{c}\small89 \\ {\tiny $(5,61)$}\end{tabular}&
\tiny $\begin {array} {cccc}
 1 & 0 & 0 & 0 \\
 0 & 1 & 0 & 0 \\
 0 & 0 & 1 & 0 \\
 0 & 0 & 0 & 1 \\
 0 & - 1 & 0 & 0 \\
 -1 & 0 & 0 & 0 \\
 -1 & 1 & 0 & 0 \\
 1 & 0 & 0 & - 1 \\
         1 & 0 & - 1 & 0
\end {array} $
&\begin{tabular}{c}\small 90 \\ {\tiny $(5,60)$}\end{tabular}&
\tiny $\begin {array} {cccc}
 1 & 0 & 0 & 0 \\
 0 & 1 & 0 & 0 \\
 0 & 0 & 1 & 0 \\
 0 & 0 & 0 & 1 \\
 0 & - 1 & 0 & 0 \\
 -1 & 0 & 0 & 0 \\
 -1 & 1 & 0 & 0 \\
 1 & 0 & - 1 & 0 \\
 0 & 0 & 1 & - 1
\end {array} $
&\begin{tabular}{c}\small 91 \\ {\tiny $(5,56)$}\end{tabular}&
\tiny $\begin {array} {cccc}
 1 & 0 & 0 & 0 \\
 0 & 1 & 0 & 0 \\
 0 & 0 & 1 & 0 \\
 0 & 0 & 0 & 1 \\
 0 & - 1 & 0 & 0 \\
 -1 & 0 & 0 & 0 \\
 -1 & 0 & - 1 & 0 \\
 1 & 1 & 0 & 0 \\
 0 & - 1 & 0 & - 1
\end {array} $
& \begin{tabular}{c}\small92 \\ {\tiny $(5,57)$}\end{tabular}&
\tiny $\begin {array} {cccc}
 1 & 0 & 0 & 0 \\
 0 & 1 & 0 & 0 \\
 0 & 0 & 1 & 0 \\
 0 & 0 & 0 & 1 \\
 0 & - 1 & 0 & 0 \\
 -1 & 0 & 0 & 0 \\
 -1 & 0 & - 1 & 0 \\
 1 & 1 & 0 & 0 \\
 1 & 0 & 0 & - 1
\end {array} $\\ \hline
\begin{tabular}{c}\small93 \\ {\tiny $(5,58)$}\end{tabular}&
\tiny $\begin {array} {cccc}
 1 & 0 & 0 & 0 \\
 0 & 1 & 0 & 0 \\
 0 & 0 & 1 & 0 \\
 0 & 0 & 0 & 1 \\
 0 & - 1 & 0 & 0 \\
 -1 & 0 & 0 & 0 \\
 0 & 0 & - 1 & - 1 \\
         1 & 0 & 0 & 1 \\
 0 & 1 & 1 & 0
\end {array} $
& \begin{tabular}{c}\small94 \\ {\tiny $(5,62)$}\end{tabular}&
\tiny $\begin {array} {cccc}
 1 & 0 & 0 & 0 \\
 0 & 1 & 0 & 0 \\
 0 & 0 & 1 & 0 \\
 0 & 0 & 0 & 1 \\
 0 & - 1 & 0 & 0 \\
 -1 & 0 & 0 & 0 \\
 0 & - 1 & - 1 & 0 \\
 1 & 1 & 0 & 0 \\
 1 & 0 & 0 & - 1
\end {array} $
&\begin{tabular}{c}\small 95 \\ {\tiny $(5,60)$}\end{tabular}&
\tiny $\begin {array} {cccc}
 1 & 0 & 0 & 0 \\
 0 & 1 & 0 & 0 \\
 0 & 0 & 1 & 0 \\
 0 & 0 & 0 & 1 \\
 0 & - 1 & 0 & 0 \\
 -1 & 0 & 0 & 0 \\
 0 & - 1 & - 1 & 0 \\
 1 & 1 & 0 & 0 \\
 1 & 1 & 0 & - 1
\end {array} $
&\begin{tabular}{c}\small 96 \\ {\tiny $(5,66)$}\end{tabular}&
\tiny $\begin {array} {cccc}
 1 & 0 & 0 & 0 \\
 0 & 1 & 0 & 0 \\
 0 & 0 & 1 & 0 \\
 0 & 0 & 0 & 1 \\
 0 & - 1 & 0 & 0 \\
 -1 & 0 & 0 & 0 \\
 1 & 1 & 0 & 0 \\
 1 & 0 & - 1 & 0 \\
 0 & 0 & 1 & - 1
\end {array} $\\ \hline
\begin{tabular}{c}\small97 \\ {\tiny $(5,69)$}\end{tabular}&
\tiny $\begin {array} {cccc}
 1 & 0 & 0 & 0 \\
 0 & 1 & 0 & 0 \\
 0 & 0 & 1 & 0 \\
 0 & 0 & 0 & 1 \\
 0 & - 1 & 0 & 0 \\
 -1 & 0 & 0 & 0 \\
 1 & 1 & 0 & 0 \\
 1 & 0 & - 1 & 0 \\
 1 & 0 & 0 & - 1
\end {array} $
& \begin{tabular}{c}\small $98_N$ \\ {\tiny $(5,60)$}\end{tabular}&
\tiny $\begin {array} {cccc}
 1 & 0 & 0 & 0 \\
 0 & 1 & 0 & 0 \\
 0 & 0 & 1 & 0 \\
 0 & 0 & 0 & 1 \\
 0 & - 1 & 0 & 0 \\
 -1 & 0 & 0 & 0 \\
 1 & 1 & 0 & 0 \\
 0 & 0 & - 1 & - 1 \\
         1 & 1 & 1 & 1
\end {array} $
&\begin{tabular}{c}\small 99 \\ {\tiny $(5,64)$}\end{tabular}&
\tiny $\begin {array} {cccc}
 1 & 0 & 0 & 0 \\
 0 & 1 & 0 & 0 \\
 0 & 0 & 1 & 0 \\
 0 & 0 & 0 & 1 \\
 0 & - 1 & 0 & 0 \\
 -1 & 0 & 0 & 0 \\
 1 & 1 & 0 & 0 \\
 0 & 1 & - 1 & 0 \\
 1 & 0 & 0 & - 1
\end {array} $
&\begin{tabular}{c}\small 100 \\ {\tiny $(5,70)$}\end{tabular}&
\tiny $\begin {array} {cccc}
 1 & 0 & 0 & 0 \\
 0 & 1 & 0 & 0 \\
 0 & 0 & 1 & 0 \\
 0 & 0 & 0 & 1 \\
 0 & - 1 & 0 & 0 \\
 -1 & 0 & 0 & 0 \\
 1 & 1 & 0 & 0 \\
 1 & 1 & - 1 & 0 \\
 0 & 0 & 1 & - 1
\end {array} $\\ \hline
\end{tabular}
\end{center}}
\end{table}
\begin{table}[thb]
{\begin{center}
\begin{tabular}{|@{}c@{}|c||@{}c@{}|c||@{}c@{}|c||@{}c@{}|c|}\hline 
no. & Vertices of $\Delta^\circ$  & no. &  Vertices of $\Delta^\circ$ & no. &  Vertices of $\Delta^\circ$ &  no. & Vertices of $\Delta^\circ$ \\ \hline \hline
\begin{tabular}{c}\small101 \\ {\tiny $(5,70)$}\end{tabular}&
\tiny $\begin {array} {cccc}
 1 & 0 & 0 & 0 \\
 0 & 1 & 0 & 0 \\
 0 & 0 & 1 & 0 \\
 0 & 0 & 0 & 1 \\
 0 & - 1 & 0 & 0 \\
 -1 & 0 & 0 & 0 \\
 1 & 1 & 0 & 0 \\
 1 & 0 & - 1 & 0 \\
 1 & 1 & 0 & - 1
\end {array} $
&\begin{tabular}{c}\small 102 \\ {\tiny $(5,75)$}\end{tabular}&
\tiny $\begin {array} {cccc}
 1 & 0 & 0 & 0 \\
 0 & 1 & 0 & 0 \\
 0 & 0 & 1 & 0 \\
 0 & 0 & 0 & 1 \\
 0 & - 1 & 0 & 0 \\
 -1 & 0 & 0 & 0 \\
 1 & 1 & 0 & 0 \\
 1 & 1 & - 1 & 0 \\
 1 & 1 & 0 & - 1
\end {array} $
& \begin{tabular}{c}\small $103_N$ \\ {\tiny $(5,57)$}\end{tabular}&
\tiny $\begin {array} {cccc}
 1 & 0 & 0 & 0 \\
 0 & 1 & 0 & 0 \\
 0 & 0 & 1 & 0 \\
 0 & 0 & 0 & 1 \\
 0 & 0 & - 1 & 0 \\
 0 & - 1 & 0 & 0 \\
 -1 & 0 & 0 & 0 \\
 0 & - 1 & - 1 & 0 \\
 1 & 1 & 1 & - 1
\end {array} $
&\begin{tabular}{c}\small 104 \\ {\tiny $(5,61)$}\end{tabular}&
\tiny $\begin {array} {cccc}
 1 & 0 & 0 & 0 \\
 0 & 1 & 0 & 0 \\
 0 & 0 & 1 & 0 \\
 0 & 0 & 0 & 1 \\
 0 & 0 & - 1 & 0 \\
 0 & - 1 & 0 & 0 \\
 -1 & 0 & 0 & 0 \\
 0 & 1 & 1 & 0 \\
 1 & 0 & 0 & - 1
\end {array} $\\ \hline
\begin{tabular}{c}\small105 \\ {\tiny $(5,63)$}\end{tabular}&
\tiny $\begin {array} {cccc}
 1 & 0 & 0 & 0 \\
 0 & 1 & 0 & 0 \\
 0 & 0 & 1 & 0 \\
 0 & 0 & 0 & 1 \\
 0 & 0 & - 1 & 0 \\
 0 & - 1 & 0 & 0 \\
 -1 & 0 & 0 & 0 \\
 1 & 1 & 0 & 0 \\
 1 & 0 & 0 & - 1
\end {array} $
& \begin{tabular}{c}\small106 \\ {\tiny $(5,59)$}\end{tabular}&
\tiny $\begin {array} {cccc}
 1 & 0 & 0 & 0 \\
 0 & 1 & 0 & 0 \\
 0 & 0 & 1 & 0 \\
 0 & 0 & 0 & 1 \\
 0 & 0 & - 1 & 0 \\
 0 & - 1 & 0 & 0 \\
 -1 & 0 & 0 & 0 \\
 1 & 1 & 0 & 0 \\
 0 & - 1 & 0 & - 1
\end {array} $
&\begin{tabular}{c}\small 107 \\ {\tiny $(5,65)$}\end{tabular}&
\tiny $\begin {array} {cccc}
 1 & 0 & 0 & 0 \\
 0 & 1 & 0 & 0 \\
 0 & 0 & 1 & 0 \\
 0 & 0 & 0 & 1 \\
 0 & 0 & - 1 & 0 \\
 0 & - 1 & 0 & 0 \\
 -1 & 0 & 0 & 0 \\
 1 & 1 & 0 & 0 \\
 1 & 1 & 0 & - 1
\end {array} $
&\begin{tabular}{c}\small $108_N$ \\ {\tiny $(5,59)$}\end{tabular}&
\tiny $\begin {array} {cccc}
 1 & 0 & 0 & 0 \\
 0 & 1 & 0 & 0 \\
 0 & 0 & 1 & 0 \\
 0 & 0 & 0 & 1 \\
 1 & - 1 & 0 & 0 \\
 -1 & 1 & 0 & 0 \\
 0 & - 1 & 0 & 0 \\
 -1 & 0 & 0 & 0 \\
 0 & 0 & - 1 & - 1
\end {array} $\\ \hline
\begin{tabular}{c}\small $109_N$ \\ {\tiny $(5,61)$}\end{tabular}&
\tiny $\begin {array} {cccc}
 1 & 0 & 0 & 0 \\
 0 & 1 & 0 & 0 \\
 0 & 0 & 1 & 0 \\
 0 & 0 & 0 & 1 \\
 1 & - 1 & 0 & 0 \\
 -1 & 1 & 0 & 0 \\
 0 & - 1 & 0 & 0 \\
 -1 & 0 & 0 & 0 \\
 1 & 0 & - 1 & - 1
\end {array} $
&\begin{tabular}{c}\small $110_N$ \\ {\tiny $(5,65)$}\end{tabular}&
\tiny $\begin {array} {cccc}
 1 & 0 & 0 & 0 \\
 0 & 1 & 0 & 0 \\
 0 & 0 & 1 & 0 \\
 0 & 0 & 0 & 1 \\
 1 & - 1 & 0 & 0 \\
 -1 & 1 & 0 & 0 \\
 0 & - 1 & 0 & 0 \\
 -1 & 0 & 0 & 0 \\
 1 & 1 & - 1 & - 1
\end {array} $
&\begin{tabular}{c}\small $111_N$ \\ {\tiny $(5,50)$}\end{tabular}&
\tiny $\begin {array} {cccc}
 1 & 0 & 0 & 0 \\
 0 & 1 & 0 & 0 \\
 0 & 0 & 1 & 0 \\
 0 & 0 & 0 & 1 \\
 1 & - 1 & 0 & 0 \\
 -1 & 1 & - 1 & 0 \\
 0 & - 1 & 0 & - 1 \\
         0 & - 1 & 1 & - 1 \\
 -1 & 1 & - 1 & 1
\end {array} $
&\begin{tabular}{c}\small $112_N$ \\ {\tiny $(5,67)$}\end{tabular}&
\tiny $\begin {array} {cccc}
 1 & 0 & 0 & 0 \\
 0 & 1 & 0 & 0 \\
 0 & 0 & 1 & 0 \\
 0 & 0 & 0 & 1 \\
 1 & - 1 & 0 & 0 \\
 -1 & 1 & 0 & 0 \\
 0 & - 1 & 0 & 0 \\
 -1 & 0 & 0 & 0 \\
 2 & 0 & - 1 & - 1
\end {array} $\\ \hline
\begin{tabular}{c}\small113 \\ {\tiny $(6,55)$}\end{tabular}&
\tiny $\begin {array} {cccc}
 1 & 0 & 0 & 0 \\
 0 & 1 & 0 & 0 \\
 0 & 0 & 1 & 0 \\
 0 & 0 & 0 & 1 \\
 0 & 0 & 0 & - 1 \\
         0 & 0 & - 1 & 0 \\
 0 & - 1 & 0 & 0 \\
 -1 & 0 & 0 & 0 \\
 1 & 1 & 0 & 0 \\
 0 & 0 & 1 & 1
\end {array} $
&\begin{tabular}{c}\small $114_N$ \\ {\tiny $(6,54)$}\end{tabular}&
\tiny $\begin {array} {cccc}
 1 & 0 & 0 & 0 \\
 0 & 1 & 0 & 0 \\
 0 & 0 & 1 & 0 \\
 0 & 0 & 0 & 1 \\
 0 & 0 & 1 & - 1 \\
         0 & 0 & - 1 & 1 \\
 0 & 0 & 0 & - 1 \\
         0 & 0 & - 1 & 0 \\
 0 & - 1 & 0 & 0 \\
 -1 & 0 & 0 & 0
\end {array} $
&\begin{tabular}{c}\small $115_N$ \\ {\tiny $(6,46)$}\end{tabular}&
\tiny $\begin {array} {cccc}
 1 & 0 & 0 & 0 \\
 0 & 1 & 0 & 0 \\
 0 & 0 & 1 & 0 \\
 0 & 0 & 0 & 1 \\
 1 & 1 & - 1 & - 1 \\
 -1 & - 1 & 1 & 1 \\
 0 & 0 & 0 & - 1 \\
         0 & 0 & - 1 & 0 \\
 0 & - 1 & 0 & 0 \\
 -1 & 0 & 0 & 0
\end {array} $
&\begin{tabular}{c}\small $116_N$ \\ {\tiny $(6,54)$}\end{tabular}&
\tiny $\begin {array} {cccc}
 1 & 0 & 0 & 0 \\
 0 & 1 & 0 & 0 \\
 0 & 0 & 1 & 0 \\
 0 & 0 & 0 & 1 \\
 0 & 1 & - 1 & 0 \\
 0 & - 1 & 1 & 0 \\
 0 & 0 & - 1 & 0 \\
 0 & - 1 & 0 & 0 \\
 -1 & 0 & 0 & 0 \\
 1 & 0 & 0 & - 1
\end {array} $\\ \hline
\begin{tabular}{c}\small $117_N$ \\ {\tiny $(6,54)$}\end{tabular}&
\tiny $\begin {array} {cccc}
 1 & 0 & 0 & 0 \\
 0 & 1 & 0 & 0 \\
 0 & 0 & 1 & 0 \\
 0 & 0 & 0 & 1 \\
 1 & 0 & - 1 & 0 \\
 -1 & 0 & 1 & 0 \\
 0 & 0 & - 1 & 0 \\
 0 & - 1 & 0 & 0 \\
 -1 & 0 & 0 & 0 \\
 1 & 0 & 0 & - 1
\end {array} $
&\begin{tabular}{c}\small $118_N$ \\ {\tiny $(6,50)$}\end{tabular}&
\tiny $\begin {array} {cccc}
 1 & 0 & 0 & 0 \\
 0 & 1 & 0 & 0 \\
 0 & 0 & 1 & 0 \\
 0 & 0 & 0 & 1 \\
 1 & - 1 & 0 & 0 \\
 -1 & 1 & 0 & 0 \\
 0 & - 1 & 0 & 0 \\
 -1 & 0 & 0 & 0 \\
 -1 & 0 & - 1 & 0 \\
 1 & 0 & 0 & - 1
\end {array} $
&\begin{tabular}{c}\small $119_N$ \\ {\tiny $(6,52)$}\end{tabular}&
\tiny $\begin {array} {cccc}
 1 & 0 & 0 & 0 \\
 0 & 1 & 0 & 0 \\
 0 & 0 & 1 & 0 \\
 0 & 0 & 0 & 1 \\
 1 & - 1 & 0 & 0 \\
 -1 & 1 & 0 & 0 \\
 0 & - 1 & 0 & 0 \\
 -1 & 0 & 0 & 0 \\
 -1 & 1 & - 1 & 0 \\
 1 & 0 & 0 & - 1
\end {array} $
& \begin{tabular}{c}\small$120_N$ \\ {\tiny $(6,56)$}\end{tabular}&
\tiny $\begin {array} {cccc}
 1 & 0 & 0 & 0 \\
 0 & 1 & 0 & 0 \\
 0 & 0 & 1 & 0 \\
 0 & 0 & 0 & 1 \\
 1 & - 1 & 0 & 0 \\
 -1 & 1 & 0 & 0 \\
 0 & - 1 & 0 & 0 \\
 -1 & 0 & 0 & 0 \\
 0 & 1 & 0 & - 1 \\
         1 & 0 & - 1 & 0
\end {array} $\\ \hline
\begin{tabular}{c}\small$121_N$ \\ {\tiny $(6,58)$}\end{tabular}&
\tiny $\begin {array} {cccc}
 1 & 0 & 0 & 0 \\
 0 & 1 & 0 & 0 \\
 0 & 0 & 1 & 0 \\
 0 & 0 & 0 & 1 \\
 1 & - 1 & 0 & 0 \\
 -1 & 1 & 0 & 0 \\
 0 & - 1 & 0 & 0 \\
 -1 & 0 & 0 & 0 \\
 1 & 0 & 0 & - 1 \\
         1 & 0 & - 1 & 0
\end {array} $
&\begin{tabular}{c}\small $122_N$ \\ {\tiny $(6,56)$}\end{tabular}&
\tiny $\begin {array} {cccc}
 1 & 0 & 0 & 0 \\
 0 & 1 & 0 & 0 \\
 0 & 0 & 1 & 0 \\
 0 & 0 & 0 & 1 \\
 1 & - 1 & 0 & 0 \\
 -1 & 1 & 0 & 0 \\
 0 & - 1 & 0 & 0 \\
 -1 & 0 & 0 & 0 \\
 1 & 0 & - 1 & 0 \\
 0 & 0 & 1 & - 1
\end {array} $
&\begin{tabular}{c}\small $123_N$ \\ {\tiny $(7,49)$}\end{tabular}&
\tiny $\begin {array} {cccc}
 1 & 0 & 0 & 0 \\
 0 & 1 & 0 & 0 \\
 0 & 0 & 1 & 0 \\
 0 & 0 & 0 & 1 \\
 0 & 0 & 1 & - 1 \\
         0 & 0 & - 1 & 1 \\
 0 & 0 & 0 & - 1 \\
         0 & 0 & - 1 & 0 \\
 0 & - 1 & 0 & 0 \\
 -1 & 0 & 0 & 0 \\
 1 & 1 & 0 & 0
\end {array} $
&\begin{tabular}{c}\small $124_N$ \\ {\tiny $(8,44)$}\end{tabular}& 
  \tiny $\begin {array} {cccc}
  1 & 0 & 0 & 0 \\
  0 & 1 & 0 & 0 \\
  0 & 0 & 1 & 0 \\
  0 & 0 & 0 & 1 \\
  1 & 0 & 0 & - 1 \\
  0 & 1 & - 1 & 0 \\
  0 & - 1 & 1 & 0 \\
  - 1 & 0 & 0 & 1 \\
  0 & 0 & 0 & - 1 \\
  0 & 0 & - 1 & 0 \\
  0 & - 1 & 0 & 0 \\
 - 1 & 0 & 0 & 0
\end {array} $\\ \hline 
\end{tabular}
\end{center}}
{\caption{\label{tb7} \sf List of the lattice vertices of $\Delta^\circ$, for each of the 124 smooth toric ambient spaces. The subscripts $P$ and $N$ indicate a product of projective spaces and a non-simple space, respectively. Each pair of integers below the space numbers denotes the two hodge numbers $h^{1,1}$ and $h^{2,1}$ of the Calabi-Yau hypersurface.}}
\end{table}

\begin{table}[thb]
{\begin{center}
\begin{tabular}{|c|c||c|c||c|c|}\hline 
no. & $~~\text{Ambient space } \mathcal A~~$ &no. & $~~\text{Ambient space } \mathcal A~~$ &no. & $~~\text{Ambient space } \mathcal A~~$ \\ \hline \hline
\begin{tabular}{c}\small $1_P$\\ {\tiny $(1,101)$}\end{tabular}& $\IP^4$ &
\begin{tabular}{c}\small $2_P$ \\ {\tiny $(2,86)$}\end{tabular}& $\IP^1 \times \IP^3$ & 
\begin{tabular}{c}\small $7_P$ \\ {\tiny $(2,83)$}\end{tabular}& $\IP^2 \times \IP^2$ \\ \hline
\begin{tabular}{c}\small17 \\ {\tiny $(3,75)$}\end{tabular}& $\IP^2 \times dP_1$ &
\begin{tabular}{c}\small $26_P$ \\ {\tiny $(3,75)$}\end{tabular}& $\IP^1 \times \IP^1 \times \IP^2$ &
\begin{tabular}{c}\small $40_P$ \\ {\tiny $(4,68)$}\end{tabular}& $\IP^1 \times \IP^1 \times \IP^1 \times \IP^1$ \\ \hline
\begin{tabular}{c}\small69 \\ {\tiny $(4,68)$}\end{tabular}& $dP_1 \times dP_1$ &
\begin{tabular}{c}\small 75 \\ {\tiny $(4,67)$}\end{tabular}& $\IP^2 \times dP_2$ &
\begin{tabular}{c}\small 78 \\ {\tiny $(4,68)$}\end{tabular}& $\IP^1 \times \IP^1 \times dP_1$ \\ \hline
\begin{tabular}{c}\small86 \\ {\tiny $(5,61)$}\end{tabular}& $\IP^1\times \IP^1 \times dP_2$ &
\begin{tabular}{c}\small 104 \\ {\tiny $(5,61)$}\end{tabular}& $dP_1 \times dP_2$ &
\begin{tabular}{c}\small $108_N$ \\ {\tiny $(5,59)$}\end{tabular}& $\IP^2 \times dP_3$ \\ \hline
\begin{tabular}{c}\small113 \\ {\tiny $(6,55)$}\end{tabular}& $dP_2 \times dP_2$ &
\begin{tabular}{c}\small $114_N$ \\ {\tiny $(6,54)$}\end{tabular}& $\IP^1 \times \IP^1 \times dP_3$ &
\begin{tabular}{c}\small $116_N$ \\ {\tiny $(6,54)$}\end{tabular}& $dP_1 \times dP_3$ \\ \hline
\begin{tabular}{c}\small $123_N$ \\ {\tiny $(7,49)$}\end{tabular}& $dP_2 \times dP_3$ &
\begin{tabular}{c}\small $124_N$ \\ {\tiny $(8,44)$}\end{tabular}& $dP_3 \times dP_3$ & & \\ \hline
\end{tabular}
\end{center}}
{\caption{\label{tb8} \sf List of the ambient spaces $\mathcal A$ which are products of del Pezzo surfaces and projective spaces. Note that only three del Pezzo surfaces $dP_{k=1,2,3}$ are toric 2-folds, and that we indeed have all the possible 17 combinations within the database of the 124. The subscripts $P$ and $N$ indicate a product of projective spaces and a non-simple space, respectively. Each pair of integers below the space numbers denotes the two hodge numbers $h^{1,1}$ and $h^{2,1}$ of the Calabi-Yau hypersurface.
}}
\end{table}

\end{appendix}


\end{document}